\documentclass[envcountsame,runningheads,a4paper]{llncs}

\usepackage{microtype}
\usepackage{mathtools,amsthm}
\usepackage[hidelinks,bookmarks,bookmarksopen,bookmarksdepth=3]{hyperref}
\hypersetup{colorlinks=false}

\usepackage{ifthen}

\newboolean{talk}
\setboolean{talk}{false}
\newboolean{paper}
\setboolean{paper}{false}

\newboolean{IEEEstyle}
\setboolean{IEEEstyle}{false}
\newboolean{lipicsstyle}
\setboolean{lipicsstyle}{false}
\newboolean{eptcsstyle}
\setboolean{eptcsstyle}{false}
\newboolean{entcsstyle}
\setboolean{entcsstyle}{false}
\newboolean{sigplanstyle}
\setboolean{sigplanstyle}{false}
\newboolean{easychairstyle}
\setboolean{easychairstyle}{false}

\newboolean{lncsstyle}
\setboolean{lncsstyle}{false}

\newboolean{needstheorems}
\setboolean{needstheorems}{false}
\newboolean{withimages}
\setboolean{withimages}{false}
\newboolean{withproofs}
\setboolean{withproofs}{true}

\newboolean{french}
\setboolean{french}{false}

\setboolean{lncsstyle}{true}
\ifthenelse{\boolean{talk}}{}{\usepackage[utf8]{inputenc}}
\ifthenelse{\boolean{french}}{\usepackage[french]{babel}}{
  \ifthenelse{\boolean{lipicsstyle}}{}{\usepackage[english]{babel}}}
\usepackage{amssymb}
\usepackage{amsmath}
\usepackage{graphicx}
\usepackage{bussproofs}
\usepackage{cmll}
\usepackage{stmaryrd}
 \usepackage{multirow}
 \ifthenelse{\boolean{talk}}{\usepackage{color}}{
   \ifthenelse{\boolean{lipicsstyle}}{\usepackage{color}}{\usepackage[usenames]{xcolor}}}
 \ifthenelse{\boolean{IEEEstyle}}{
 \usepackage[bookmarks={false}]{hyperref}}{
 \usepackage{hyperref}}
 \usepackage{fancybox}
 \usepackage{xspace}
 \usepackage{hhline}
\ifthenelse{\boolean{eptcsstyle}}{}{
	\ifthenelse{\boolean{entcsstyle}}{}{
\usepackage{txfonts}}}
\usepackage{nameref}
\usepackage{appendix}

\ifthenelse{\boolean{IEEEstyle}}{
\setboolean{needstheorems}{true}}{}

\ifthenelse{\boolean{eptcsstyle}}{
\setboolean{needstheorems}{true}}{}

\ifthenelse{\boolean{sigplanstyle}}{
\setboolean{needstheorems}{true}}{}

\ifthenelse{\boolean{easychairstyle}}{
\setboolean{needstheorems}{true}}{}

\ifthenelse{\boolean{lipicsstyle}}{
    \newtheorem{proposition}[theorem]{Proposition}
    }{}

\ifthenelse{\boolean{needstheorems}}{

\usepackage{amsthm}

    \newtheorem{theorem}{Theorem}[section]
    \newtheorem{lemma}[theorem]{Lemma}
    \newtheorem{corollary}[theorem]{Corollary}
    \newtheorem{proposition}[theorem]{Proposition}
    \newtheorem{definition}[theorem]{Definition}
    \newtheorem{remark}[theorem]{Remark}
}{}

\newcommand{\myproof}[1]{
\ifthenelse{\boolean{withproofs}}{#1}{}
}

\newcommand{\la}[1]{\lambda #1.}

\newcommand{\tm}{t}
\newcommand{\tmtwo}{s}
\newcommand{\tmthree}{u}
\newcommand{\tmfour}{r}
\newcommand{\tmfive}{p}

\newcommand{\var}{x}
\newcommand{\vartwo}{y}
\newcommand{\varthree}{z}
\newcommand{\varfour}{w}

\newcommand{\varset}{\Delta}

\newcommand{\nameset}{\Delta}
\newcommand{\namesettwo}{\Gamma}
\newcommand{\namesetthree}{\Pi}

\newcommand{\Rew}[1]{\rightarrow_{#1}}

\renewcommand{\to}{\Rew{}}

\newcommand{\gcsym}{{\tt gc}}
\newcommand{\gc}{\gcsym}

\newcommand{\esym}{{\mathtt e}}
\newcommand{\wsym}{{\mathtt w}}

\newcommand{\msym}{{\mathtt m}}

\newcommand{\ctxholep}[1]{\langle #1\rangle}
\newcommand{\ctxhole}{\ctxholep{\cdot}}
\newcommand{\ctxholen}[1]{\ctxholep{\cdot}_{#1}}
\newcommand{\ctxholens}{\ctxholen{\nameset}}
\newcommand{\ctxholenp}[2]{\ctxholep{#2}_{#1}}
\newcommand{\ctxholensp}[1]{\ctxholenp{\nameset}{#1}}

\newcommand{\ctxholefp}[1]{\langle #1\rangle}

\makeatletter
\newsavebox{\@brx}
\newcommand{\llangle}[1][]{\savebox{\@brx}{\(\m@th{#1\langle}\)}%
  \mathopen{\copy\@brx\kern-0.7\wd\@brx\usebox{\@brx}}}
\newcommand{\rrangle}[1][]{\savebox{\@brx}{\(\m@th{#1\rangle}\)}%
  \mathclose{\copy\@brx\kern-0.7\wd\@brx\usebox{\@brx}}}
\makeatother

\renewcommand{\ctxholefp}[1]{\llangle #1\rrangle}

\newcommand{\ctx}{C}
\newcommand{\ctxtwo}{D}
\newcommand{\ctxthree}{E}

\newcommand{\ctxp}[1]{\ctx\ctxholep{#1}}

\newcommand{\ctxfp}[1]{\ctx\ctxholefp{#1}}

\newcommand{\ctxn}[1]{\ctx_{#1}}
\newcommand{\ctxns}{\ctxn\nameset}

\newcommand{\ctxnp}[2]{\ctxn{#1}\ctxholep{#2}}
\newcommand{\ctxnsp}[1]{\ctxns\ctxholep{#1}}

\newcommand{\nbvctxtwo}[1]{\nbvctxtwo{#1}}

\newcommand{\sctx}{L}
\newcommand{\sctxtwo}{{L'}}
\newcommand{\sctxthree}{{L''}}
\newcommand{\sctxp}[1]{\sctx\ctxholep{#1}}

\newcommand{\defeq}{:=}
\newcommand{\grameq}{::=}

\newcommand{\isub}[2]{\{#1/#2\}}
\newcommand{\esub}[2]{[#1/#2]}
\renewcommand{\esub}[2]{[#1{\shortleftarrow}#2]}
\renewcommand{\isub}[2]{\{#1{\shortleftarrow}#2\}}

\newcommand{\letexp}[3]{{\tt let}\ #1=#2\ {\tt in}\ #3}

\newcommand{\todb}{\todbp{}}

\newcommand{\tols}{\Rew{\lssym}}
\newcommand{\togc}{\Rew{\gcsym}}

\newcommand{\llbrace}{\{ \kern -0.27em \vert}
\newcommand{\rrbrace}{\vert \kern -0.27em \}}

\renewcommand{\l}{\lambda}
\newcommand{\ie}{{\em i.e.}\xspace}
\newcommand{\eg}{{\em e.g.}\xspace}
\newcommand{\ih}{{\textit{i.h.}}\xspace}
\newcommand{\fv}[1]{{\tt fv}(#1)}

\ifthenelse{\boolean{entcsstyle}}{}{
	\newcommand{\cv}[1]{{\tt cv}(#1)}}

\newcommand{\red}[1]{{\color{red} {#1}}}

\newcommand{\ignore}[1]{}

\newcommand{\colspace}{@{\hspace{.5cm}}}
\newcommand{\myinput}[1]{\ifthenelse{\boolean{withimages}}{\input{#1}}{}}
\newcommand{\spc}{@{\hspace{.35cm}}}

\newcommand{\reflemma}[1]{Lemma~\ref{l:#1}}

\newcommand{\refthm}[1]{Theorem~\ref{thm:#1}}

\newcommand{\refprop}[1]{Proposition~\ref{prop:#1}}

\ifthenelse{\boolean{easychairstyle}}{}{
  \ifthenelse{\boolean{lncsstyle}}{}{	
	
  }
}
\newcommand{\reffig}[1]{Fig.~\ref{fig:#1}}

\newcommand{\mellies}{{Melli{\`e}s}\xspace}
\newcommand{\levy}{{L{\'e}vy}\xspace}

\newcommand{\form}{A}
\newcommand{\formtwo}{B}

\newcommand{\tens}{\otimes}

\newcommand{\bang}{{\mathsf{!}}}

\newcommand{\der}{{\mathsf{d}}}

\newcommand{\weak}{{\mathsf{w}}}

\newcommand{\set}[1]{\{#1\}}

\newcommand{\nat}{\mathbb{N}}

\newcommand{\zeronet}[1]{#1^0}

\newcommand{\bbox}{{\tt ibox}}
\newcommand{\bboxp}[1]{\bbox(#1)}

\newcommand{\genax}{\tt genax}

\newcommand{\link}{l}
\newcommand{\linktwo}{h}
\newcommand{\linkthree}{j}

\newcommand{\net}{P}
\newcommand{\nettwo}{Q}
\newcommand{\netthree}{R}
\newcommand{\netfour}{S}

\newcommand{\tolsc}{\Rew{LSC}}

\newcommand{\tom}{\Rew{\msym}}
\newcommand{\toe}{\Rew{\esym}}
\newcommand{\tow}{\Rew{\wsym}}

\newcommand{\indproof}[2]{
 	\AxiomC{#2}
 	\noLine
 	\UnaryInfC{\tiny :}
  	\noLine
 	\UnaryInfC{#1}
 }

\newcommand{\lamtonets}[1]{\underline{#1}}
\newcommand{\lamtonetsvar}[2]{\underline{#1}_{#2}}

\newcommand{\sizep}[2]{|#1|_{#2}}

\newcommand{\imagespath}{images}

\newcommand{\eqstruct}{\equiv}

\newcommand{\withproofs}[1]{\ifthenelse{\boolean{withproofs}}{#1}{}}

\newcommand{\withoutproofs}[1]{\ifthenelse{\boolean{withproofs}}{}{#1}}

\ifthenelse{\boolean{entcsstyle}}{}{
  \ifthenelse{\boolean{lncsstyle}}{}{	
    
  }
 }

\newcounter{numberone}

\newcommand{\med}[2]{
($(#1)!.5!(#2)$)
}

\newcommand{\nodeset}[1]{{\tt nodes}(#1)}
\newcommand{\linkset}[1]{{\tt links}(#1)}

\newcommand{\readback}{\rhd}

\newcommand{\expr}{e}
\newcommand{\exprtwo}{f}
\newcommand{\exprthree}{g}
\newcommand{\exprfour}{h}

\newcommand{\redex}{R}

\usepackage{tikz}
\usetikzlibrary{calc}
\usetikzlibrary{matrix,arrows}
\usetikzlibrary{decorations.shapes}
\usetikzlibrary{decorations.text}
\usetikzlibrary{decorations.pathmorphing}
\usetikzlibrary{decorations.markings}
\usetikzlibrary{positioning}

\tikzset{
node distance=1.3cm, auto,
every node/.style={font=\scriptsize },
ocenter/.style={baseline={([yshift=-.5ex, xshift=-.5ex]current bounding box)}},  
labelBeginAbove/.style={postaction={decorate,decoration={markings,mark=at position 0 with {\node[inner sep= 0.6pt, above=1pt]{\tiny #1};}} } },
labelBeginBelow/.style={postaction={decorate,decoration={markings,mark=at position 0 with {\node[inner sep= 0.6pt, below=1pt]{\tiny #1};}}}},
labelEndAbove/.style={postaction={decorate,decoration={markings,mark=at position 1 with {\node[inner sep= 0.6pt, above=1pt]{\tiny #1};}}}},
labelEndBelow/.style={postaction={decorate,decoration={markings,mark=at position 1 with {\node[inner sep= 0.6pt, below=1pt]{\tiny #1};}}}},
labelEndRight/.style={postaction={decorate,decoration={markings,mark=at position 1 with {\node[inner sep= 0.6pt, right=1pt]{\tiny #1};}}}},
labelEndLeft/.style={postaction={decorate,decoration={markings,mark=at position 1 with {\node[inner sep= 0.6pt, left=1pt]{\tiny #1};}}}}
}

\usepackage{tikz}
\usetikzlibrary{calc}
\usetikzlibrary{backgrounds}
\usetikzlibrary{decorations.shapes}
\usetikzlibrary{decorations.text}
\usetikzlibrary{decorations.pathmorphing}
\usetikzlibrary{decorations.markings}
\usetikzlibrary{matrix}
\usetikzlibrary{decorations.pathreplacing}
\usetikzlibrary{arrows}
\usetikzlibrary{positioning}
\usepackage[nofancy]{tikz-inet}

\setboolean{withimages}{true}

\tikzset{%
  ocenter/.style={baseline={([yshift=-.5ex, xshift=-.5ex]current bounding box)}},  
  nospace/.style={inner sep= 0pt},
  etic/.style={inner sep= 0.5pt, fill=white, anchor= center},
  letic/.style={inner sep= 0.7pt, fill=white, anchor= center,circle,draw=black},
  port/.style={inner sep= 0.3pt, anchor= center,circle,draw=black,fill=black, minimum size = 0.3pt},
  eport/.style={inner sep= 0.8pt, anchor= center,circle,draw=cyan,fill=white, minimum size = 0.8pt},
  mport/.style={inner sep= 0.8pt, anchor= center, circle,draw=white,fill=brown, minimum size = 0.8pt, solid, 
line width=0.10ex},
  eportn/.style={inner sep= 0.8pt, anchor= center,circle,draw=blue,fill=white, minimum size = 0.8pt},
  mportn/.style={inner sep= 0.8pt, anchor= center, circle,draw=red,fill=red, minimum size = 0.8pt, solid, 
line width=0.10ex},
  nopol/.style={->, shorten <=0.5pt, shorten >=0.5pt, draw=gray, line width=0.18ex},
  nopolrev/.style={<-, shorten <=1pt, shorten >=1pt, draw=gray, line width=0.18ex},
  nopolgen/.style={preaction={decorate},decoration={markings,mark=at position .5 with {\draw [shorten >=0pt, shorten <=0pt,draw=gray,-](0pt,3pt) -- (0pt,-3pt);}},->, shorten <=0.5pt, shorten >=0.5pt, draw=gray, line width=0.18ex},
  nopolrevgen/.style={preaction={decorate},decoration={markings,mark=at position .5 with {\draw [shorten >=0pt, shorten <=0pt,draw=gray,-](0pt,3pt) -- (0pt,-3pt);}},<-, shorten <=1pt, shorten >=1pt, draw=gray, line width=0.18ex},
  outEdge/.style={<-, shorten >=0.5pt, shorten <=0.5pt, draw=blue, line width=0.18ex},
  inpEdge/.style={->, densely dotted, shorten >=0.5pt, shorten <=0.5pt, draw=red, line width=0.18ex, overlay},
  inpEdgeGen/.style={preaction={decorate},decoration={markings,mark=at position .5 with {\draw [shorten >=0pt, shorten <=0pt,draw=red,-](0pt,3pt) -- (0pt,-3pt);}},->, densely dotted, shorten >=0.5pt, shorten <=0.5pt, draw=red, line width=0.18ex, overlay},
    eprinc/.style={postaction={decorate},decoration={markings,mark=at position .4 with {\node [inner sep= 0.5pt, anchor= center,circle,draw=cyan, fill=cyan, minimum size = 0.8pt, solid, line width=0.10ex]{};}}},
    mprinc/.style={postaction={decorate},decoration={markings,mark=at position .4 with {\node [inner sep= 0.5pt, anchor= center,circle,draw=brown, fill=brown, minimum size = 0.8pt, solid, line width=0.10ex]{};}}},
    mprincpar/.style={postaction={decorate},decoration={markings,mark=at position .6 with {\node [inner sep= 0.5pt, anchor= center,circle,draw=brown, fill=brown, minimum size = 0.8pt, solid, line width=0.10ex]{};}}},
    pure/.style={->, shorten <=1pt, shorten >=1pt, draw=brown, line width=0.18ex},
    pureRev/.style={<-, shorten <=1pt, shorten >=1pt, draw=brown, line width=0.18ex},
    exppure/.style={->, shorten <=1pt, shorten >=1pt, draw=cyan, line width=0.18ex, densely dotted},
    exppureRev/.style={<-, shorten <=1pt, shorten >=1pt, draw=cyan, line width=0.18ex, densely dotted},    genexppure/.style={preaction={decorate},decoration={markings,mark=at position .5 with {\draw [shorten >=0pt, shorten <=0pt,draw=cyan,-](0pt,3pt) -- (0pt,-3pt);}},->, shorten <=1pt, shorten >=1pt, draw=cyan, line width=0.18ex, densely dotted},
    enaryedge/.style={preaction={decorate},decoration={markings,mark=at position .5 with {\draw [shorten >=0pt, shorten <=0pt,draw=cyan,-](0pt,3pt) -- (0pt,-3pt);}}},
    eprincn/.style={postaction={decorate},decoration={markings,mark=at position .4 with {\node [inner sep= 0.7pt, anchor= center,circle,draw=blue, fill=blue, minimum size = 0.8pt, solid, line width=0.10ex]{};}}},
    mprincn/.style={postaction={decorate},decoration={markings,mark=at position .6 with {\node [inner sep= 0.7pt, anchor= center,circle,draw=red, fill=white, minimum size = 0.8pt, solid, line width=0.10ex]{};}}},
    mprincnpar/.style={postaction={decorate},decoration={markings,mark=at position .4 with {\node [inner sep= 0.7pt, anchor= center,circle,draw=red, fill=white, minimum size = 0.8pt, solid, line width=0.10ex]{};}}},
    puren/.style={->, shorten <=1pt, shorten >=1pt, draw=red, line width=0.18ex},
    purenRev/.style={<-, shorten <=1pt, shorten >=1pt, draw=red, line width=0.18ex},
    exppuren/.style={->, shorten <=1pt, shorten >=1pt, draw=blue, line width=0.18ex, densely dotted},
    exppurenRev/.style={<-, shorten <=1pt, shorten >=1pt, draw=blue, line width=0.18ex, densely dotted},    genexppuren/.style={preaction={decorate},decoration={markings,mark=at position .5 with {\draw [shorten >=0pt, shorten <=0pt,draw=blue,-](0pt,3pt) -- (0pt,-3pt);}},->, shorten <=1pt, shorten >=1pt, draw=blue, line width=0.18ex, densely dotted},
    enaryedge/.style={preaction={decorate},decoration={markings,mark=at position .5 with {\draw [shorten >=0pt, shorten <=0pt,draw=blue,-](0pt,3pt) -- (0pt,-3pt);}}},
  jboxline/.style={draw= gray,rounded corners, line width=0.20ex, overlay},
  exboxline/.style={draw= gray,line width=0.20ex, overlay},
  noboxline/.style={draw= white,rounded corners, line width=0ex},
  net/.style={draw=gray,inner sep=2pt,thick,ellipse, anchor=center, font=\scriptsize},
  inductiveTr/.style={draw=black!50, minimum size=0.9cm},
  inductiveTrSmall/.style={draw=black!50, minimum size=0.6cm},
  every label/.style={label distance = 1pt, font=\scriptsize, inner sep= 1pt},  
  every node/.style={font=\tiny }
}

\newcommand{\altax}{8pt}

\newcommand{\stalt}{22pt}
\newcommand{\stlar}{15pt}
\newcommand{\hstalt}{\stalt/2}
\newcommand{\hstlar}{\stlar/2}

\newcommand{\ilar}{12pt}

\newcommand{\lUnarySymbol}[5]{
\node at ($(#2.center) ! .5 ! (#1.center)$) [#5] (#3){ #4};
}

\newcommand{\lUnaryDiffEdgesCircleNode}[7]{

\lUnarySymbol{#1}{#2}{#3}{#4}{#7}

\draw[#5] (#3) to (#1);

\draw[#6] (#2) to (#3);
}

\newcommand{\lBinSymbol}[6]{
\node at ($(#2.center) ! .5 ! (#3.center) ! .5 ! (#1.center)$) [#6] (#4){#5};
}

\newcommand{\lBinSymbolFixR}[6]{
\node at ($(#1.center) ! .5 ! (#3 -| #1)$) [below= 1pt, #6] (#4){#5};
}

\newcommand{\lBinEdgesBelow}[5]{
\draw[#4, in=210, out=90] (#1) to (#3);
\draw[#5, in=-30, out=90] (#2) to (#3);
}

\newcommand{\lBinaryWithEdgeBelowTypes}[9]{
\lBinSymbol{#1}{#2}{#3}{#4}{#5}{#9}

\lBinEdgesBelow{#2}{#3}{#4}{#7}{#8}

\draw[#6] (#4) to (#1);
}

\newcommand{\sepbox}{2pt}

\newcommand{\boxnodes}[4]{
\node at (#1.center)[left = #2,nospace](#1so){};
\node at (#1.center)[right = #3,nospace](#1se){};
\node at (#1se.center)[above=#4,nospace](#1ne){};
\node at (#1ne-|#1so)[nospace](#1no){};}

\newcommand{\BoxNodesWithPalOnTopAndVeticalNodeBound}[4]{
\node at (#1.center)[left = #3,nospace](#1no){};
\node at (#1.center)[right = #4,nospace](#1ne){};
\node at (#1no|-#2)[nospace](#1so){};
\node at (#1ne|-#2)[nospace](#1se){};
}

\newcommand{\boxline}[2]{
\draw[#2](#1.center) -- (#1se.center) -- (#1ne.center) -- (#1no.center) -- (#1so.center)--(#1.center);}

\newcommand{\BoxLineWithPalOnTop}[2]{
\draw[#2](#1so.center) -- (#1se.center) -- (#1ne.center) -- (#1.center) -- (#1no.center) -- (#1so.center); 
}

\newcommand{\abox}[5]{
\boxnodes{#1}{#3}{#4}{#5}
\boxline{#1}{#2line}}

\newcommand{\aBoxWithPalOnTopAndVeticalNodeBound}[5]{
\BoxNodesWithPalOnTopAndVeticalNodeBound{#1}{#2}{#4}{#5}
\BoxLineWithPalOnTop{#1}{#3line}
}

\newcommand{\boxdershorizontaldistance}{7pt}
\newcommand{\boxdersverticaldistance}{2pt}
\newcommand{\boxdersdots}{5pt}

\newcommand{\boxcornerright}[4]{

\node at (#1.center)[left=#2, nospace](#1lineaso){};
\node at (#1.center)[above=#3, nospace](#1lineane){};
\draw[#4](#1lineaso.center) -- (#1.center) -- (#1lineane.center);

\node at (#1lineaso.center)[left=\boxdersdots, nospace](#1lineasodotted){};
\node at (#1lineane.center)[above=\boxdersdots, nospace](#1lineanedotted){};
\draw[#4, densely dotted](#1lineaso)to(#1lineasodotted);
\draw[#4, densely dotted](#1lineane)to(#1lineanedotted);
}

\newcommand{\boxcornersright}[4]{
\node at (#1.center)[above right=2pt and 4pt, etic](#1dots){\tiny $...$};
\boxcornerright{#1}{#2}{#3}{#4}

\node at (#1.center)[below right=\boxdersverticaldistance and \boxdershorizontaldistance, nospace](#1linease2){};
\node at (#1lineaso.center)[below=\boxdersverticaldistance, nospace](#1lineaso2){};

\node at (#1lineane.center)[right=\boxdershorizontaldistance, nospace](#1lineane2){};
\draw[#4](#1lineaso2.center) -- (#1linease2.center) -- (#1lineane2.center);

\node at (#1lineaso2.center)[left=\boxdersdots, nospace](#1lineaso2dotted){};
\node at (#1lineane2.center)[above=\boxdersdots, nospace](#1lineane2dotted){};
\draw[#4, densely dotted](#1lineaso2)to(#1lineaso2dotted);
\draw[#4, densely dotted](#1lineane2)to(#1lineane2dotted);

}

\newcommand{\lparn}[4]{
\lBinaryWithEdgeBelowTypes{#1}{#2}{#3}{#4}{{\scriptsize$\parr$}}{mprincnpar,puren}{exppuren}{puren}{below=.05cm,letic}
}

\newcommand{\lparnangle}[5]{

\lBinSymbolFixR{#1}{#2}{#3}{#4}{\scriptsize$\parr$}{letic}

\draw[exppuren, #5] (#2) to (#4);
\draw[purenRev, out=-30, in=90] (#4) to (#3);
\draw[mprincn, purenRev] (#1) to (#4);
}

\newcommand{\lparnanglefake}[5]{
\lBinSymbolFixR{#1}{#2}{#3}{#4}{}{letic}
}

\newcommand{\ltensn}[4]{
\lBinaryWithEdgeBelowTypes{#1}{#2}{#3}{#4}{\scriptsize$\tens$}{puren}{puren}{exppurenRev}{letic}
\draw[mprincn,purenRev, out=210, in=90] (#4) to (#2);
}
\newcommand{\lbangn}[3]{
\lUnaryDiffEdgesCircleNode{#1}{#2}{#3}{\scriptsize $!$}{purenRev}{eprincn, exppuren}{below=.03cm,letic}
}

\newcommand{\ldern}[3]{
\lUnaryDiffEdgesCircleNode{#1}{#2}{#3}{\scriptsize $\der$}{eprincn, exppuren}{purenRev}{above=.03cm,letic}
}

\newcommand{\ldernangle}[4]{
\node at ($(#2.center) ! .5 ! (#2 |- #1)$) [above= .5pt,fill=white, letic] (#3){\scriptsize $\der$};
\draw[puren] (#2) to (#3);
\draw[eprincn, exppuren, #4] (#3) to (#1);
}

\newcommand{\lweakn}[2]
{
 \node at (#1.center) [above= 1.1*\hstalt,fill=white, letic] (#2){\scriptsize$\weak$};

\draw[eprincn, exppuren] (#2) to (#1);

}
\newcommand{\lcolboxn}[4]{

\lBinaryWithEdgeBelowTypes{#1}{#2}{#3}{#4}{\scriptsize$\genax$}{exppurenRev}{exppurenRev}{exppurenRev}{letic}

}

\newcommand{\lcontextn}[4]{

\lBinaryWithEdgeBelowTypes{#1}{#2}{#3}{#4}{\scriptsize$\ctxhole$}{puren}{exppurenRev}{exppurenRev}{letic}

}

\newcommand{\gdots}[3]{
\node at ($(#1)!.5!(#2)$) [#3]{\tiny $\ldots$};
}

\newcommand{\gdotsname}[4]{
\node at ($(#1)!.5!(#2)$) [#3](#4){\tiny $\ldots$};
}

\newcommand{\nnode}{u}
\newcommand{\nnodetwo}{w}

\newcommand{\rootnode}{r}

\renewcommand{\path}{\tau}

\renewcommand{\genax}{\Box}

\newcommand{\sepboxshort}{1pt}
\renewcommand{\sepbox}{5pt}
\renewcommand{\stalt}{28pt}
\renewcommand{\stlar}{20pt}

\tikzset{every label/.style={label distance = 1pt, font=\footnotesize, inner sep= 1pt}}

\renewcommand{\redex}{\gamma}

 \renewcommand{\todb}{\tom}
 \renewcommand{\tols}{\toe}

\setboolean{withproofs}{true}
\setboolean{withimages}{true}

\begin{document}
\mainmatter  

\title{Proof Nets and the Linear Substitution Calculus}

\titlerunning{Proof Nets and the Linear Substitution Calculus}

%
%
\author{Beniamino Accattoli}
\authorrunning{Accattoli}


\institute{}
\institute{Inria, UMR 7161, LIX, \'Ecole Polytechnique\\ \email{\href{mailto:beniamino.accattoli@inria.fr}{beniamino.accattoli@inria.fr}}}

%
%

\toctitle{Lecture Notes in Computer Science}
\tocauthor{Authors' Instructions}
\maketitle

\begin{abstract}
Since the very beginning of the theory of linear logic it is known how to represent the $\l$-calculus as linear logic proof nets. The two systems however have different granularities, in particular proof nets have an explicit notion of sharing---the exponentials---and a micro-step operational semantics, while the $\l$-calculus has no sharing and a small-step operational semantics. Here we show that the \emph{linear substitution calculus}, a simple refinement of the $\l$-calculus with sharing, is isomorphic to proof nets at the operational level.

Nonetheless, two different terms with sharing can still have the same proof nets representation---a further result is the characterisation of the equality induced by proof nets over terms with sharing. Finally, such a detailed analysis of the relationship between terms and proof nets, suggests a new, abstract notion of proof net, based on rewriting considerations and not necessarily of a graphical nature.
\end{abstract}





\section{Introduction}
Girard's seminal paper on linear logic \cite{DBLP:journals/tcs/Girard87} showed how to represent intuitionistic logic---and so the $\l$-calculus---inside linear logic. During the nineties, Danos and Regnier provided a detailed study of such a representation via proof nets \cite{Danos:Thesis:90,Reg:Thesis:92,Danos:1995:PHS:212876.212903,DBLP:journals/tcs/DanosR99}, which is
nowadays a cornerstone of the field. 
Roughly, linear logic gives first-class status to \emph{sharing}, accounted for by the \emph{exponential} layer of the logic, and not directly visible in the $\l$-calculus. In turn, cut-elimination in linear logic provides a micro-step refinement of the small-step operational semantics of the $\l$-calculus, that is, $\beta$-reduction. 

\paragraph{The mismatch.}  Some of the insights provided by proof nets cannot be directly expressed in the $\l$-calculus, because of the mismatch of granularities. Typically, there is a \emph{mismatch of states}: simulation of $\beta$ on proofs passes through intermediate states / proofs that cannot be expressed as $\l$-terms. The mismatch does not allow, for instance, expressing fine strategies such as linear head evaluation \cite{DBLP:journals/tcs/MascariP94,Danos04headlinear} in the $\l$-calculus, nor to see in which sense proof nets quotient terms, as such a quotient concerns only the intermediate proofs. And when one starts to have a closer look, there are other mismatches, of which the lack of sharing in the $\l$-calculus is only the most macroscopic one.

Some minor issues are due to a \emph{mismatch of styles}: the fact that terms and proofs, despite their similarities, have different representations of variables and notions of redexes. Typically, two occurrences of a same variable in a term are smoothly identified by simply using the same name, while for proofs there is an explicit rule, contraction, to identify them. Name identification is obviously associative, commutative, and commutes with all constructors, while contractions do not have these properties for free\footnote{$\alpha$-equivalence is subtle on terms, but this is an orthogonal issue, and a formal approach to proof net should also deal with $\alpha$-equivalence for nodes, even if this is never done.}. For redexes, the linear logic representation of terms has many cuts with axioms that have no counterpart on terms. These points have been addressed in the literature, using for instance generalised contractions or interaction nets, but they are not devoid of further technical complications. Establishing a precise relationship between terms and proofs and their evaluations is, in fact, a very technical affair.

A serious issue is the \emph{mismatch of operational semantics}. The two systems compute the same results, but with different  rewriting rules, and linear logic is far from having the nice rewriting properties of the $\l$-calculus. Typically, the $\l$-calculus has a \emph{residual system} \cite{Terese}\footnote{For the unacquainted reader: having a residual system means to be a well-behaved rewriting system---related concepts are orthogonal systems, or the parallel moves or cube properties.}, which is a strong form of confluence that allows building its famous advanced rewriting theory, given by standardisation, neededness, and \levy's optimality \cite{thesislevy}. In the ordinary presentations of linear logic cut-elimination is confluent but it does not admit residual systems\footnote{Some presentations of proof nets (\eg Regnier's in \cite{Reg:Thesis:92}) solve the operational semantics mismatch adapting proof nets to the $\l$-calculus, and do have residuals, but then they are unable to express typical micro-step proof nets concepts such as linear head reduction.}, and so the advanced rewriting properties of the $\l$-calculus are lost. Put differently, linear logic is a structural refinement of the $\l$-calculus but it is far from refining it at the rewriting level.

A final point is the \emph{mismatch of representations}: proofs in linear logic are usually manipulated in their graphical form, that is, as proof nets, and, while this is a handy formalism for intuitions, it is not amenable to formal reasoning---it is not by chance that there is not a single result about proof nets formalised in a proof assistant. And as already pointed out, the parallelism provided by proof nets, in the case of the $\l$-calculus, shows up only in the nets obtained as intermediate steps of the simulation of $\beta$, and so it cannot easily be seen on the $\l$-calculus. There is a way of expressing it, known as $\sigma$-equivalence, due to Regnier \cite{regnier94}, but it is far from being natural.

\paragraph{The linear substitution calculus.} The linear substitution calculus (LSC) \cite{DBLP:conf/rta/Accattoli12,DBLP:conf/popl/AccattoliBKL14} is a refinement of the $\l$-calculus with sharing, introduced by Accattoli and Kesner as a minor variation over a calculus by Milner \cite{DBLP:journals/entcs/Milner07}, and meant to correct all these problems at once. 

The LSC has been introduced in 2012 and then used in different settings---a selection of relevant studies concerning cost models, standardisation, abstract machines, intersection types, call-by-need, the $\pi$-calculus, and \levy's optimality is \cite{DBLP:journals/corr/AccattoliL16,DBLP:conf/popl/AccattoliBKL14,DBLP:conf/icfp/AccattoliBM14,DBLP:conf/ifipTCS/KesnerV14,DBLP:conf/fossacs/Kesner16,DBLP:journals/corr/abs-1302-6337,DBLP:conf/rta/BarenbaumB17}. The two design features of the LSC are its tight relationship with proof nets and the fact of having a residual system. The matching with proof nets, despite being one of the two reasons to be of the LSC, for some reason was never developed in detail, nor published. This paper corrects the situation, strengthening a growing body of research. 

\paragraph{Contributions.} The main result of the paper is the perfect correspondence between the LSC and the fragment of linear logic representing the $\l$-calculus. To this goal, the presentation of proof nets has to be adjusted, because the fault for the mismatch is not always on the calculus side. To overcome the mismatch of styles, we adopt a presentation of proof nets---already at work by the author \cite{DBLP:journals/tcs/Accattoli15}---that intuitively corresponds to interaction nets (to work modulo cut with axioms) with \emph{hyper-wires}, that is, wires connecting more than two ports (to have smooth contractions). Our presentation of proof nets also refines the one in \cite{DBLP:journals/tcs/Accattoli15} with a micro-step operational semantics. Our exponential rewriting rules are slightly different than the others in the literature, and look more as the replication rule of the $\pi$-calculus---this is the key change for having a residual system.

Essentially, the LSC and our proof nets presentation are isomorphic. More precisely, our contribution is to establish the following tight correspondence:
\begin{enumerate}
  \item \emph{Transferable syntaxes}: every term translates to a proof net, and every proof net reads back to at least one term, removing the mismatch of states. We rely on a correctness criterion---Laurent's one for polarised proof nets \cite{DBLP:journals/tcs/Laurent03,phdlaurent}---to characterise proof nets and read them back. There can be many terms mapping to the same proof net, so at this level the systems are not isomorphic. 
  \item \emph{Quotient}: we characterise the simple equivalence $\eqstruct$ on terms that is induced by the translation to proof nets. The quotient of terms by $\eqstruct$ is then isomorphic to proof nets, refining the previous point. The characterisation of the quotient is not usually studied in the literature on proof nets.
  \item \emph{Isomorphic micro-step operational semantics}: a term $\tm$ and its associated proof net $\net$ have redexes in bijection, and such a bijection is a strong bisimulation: one step on one side is simulated by exactly one step on the other side, and vice-versa, and in both cases the reducts are still related by translation and read back. Therefore, the mismatch of operational semantics also vanishes. 

  The fact that the LSC has a residual system is proved in \cite{DBLP:conf/popl/AccattoliBKL14}, and it is not treated here. But our results allow to smoothly transfer the residual system from the LSC to our presentation of proof nets.
\end{enumerate}

These features allow to consider the LSC modulo $\eqstruct$ as an algebraic---that is, not graphical---reformulation of proof nets for the $\l$-calculus, providing the strongest possible solution to the mismatch of representations. At the end of the paper, we also suggest a new perspective on proof nets from a rewriting point of view, building on our approach.

\paragraph{The value of this paper.} This work is a bit more than the filling of a gap in the literature. The development is detailed, and so necessarily technical, and yet clean. The study of correctness and sequentialisation is stronger than in other works in the literature, because beyond sequentialising we also characterise the quotient---the proof of the characterisation is nonetheless pleasantly simple. Another unusual point is the use of \emph{context nets} corresponding to the contexts of the calculus, that are needed to deal with the rules of the LSC. Less technically, but maybe more importantly, the paper ends with the sketch of a new and high-level rewriting perspective on proof nets. 

\paragraph{Proofs.} \ifthenelse{\boolean{withproofs}}{For lack of space, all proofs have been moved to the Appendix.}{For lack of space, all proofs have been moved to the technical report \cite{}.}

\subsection{Historical Perspective} 

The fine match between the LSC and proof nets does not come out of the blue: it rather is the final product of a decades-long quest for a canonical decomposition of the $\l$-calculus. 

At the time of the introduction of linear logic, decompositions of the $\lambda$-calculus arose also from other contexts. Abadi, Cardelli, Curien, and \levy introduced calculi with \emph{explicit substitutions} \cite{DBLP:journals/jfp/AbadiCCL91}, that are refinements of the $\l$-calculus where meta-level substitution is delayed, by introducing explicit annotations, and then computed in a micro-step fashion.  A decomposition of a different nature appeared in concurrency, with the translations of the $\l$-calculus to the $\pi$-calculus \cite{DBLP:journals/mscs/Milner92}, due to Milner.

These settings introduce an explicit treatment of \emph{sharing}---called \emph{exponentials} in linear logic, or explicit substitutions, or \emph{replication} in the $\pi$-calculus. 
The first calculus of explicit substitutions suffered of a design issue, as showed by \mellies in \cite{DBLP:conf/tlca/Mellie95}. A turning point was the link between explicit substitutions and linear logic proof nets by Di Cosmo and Kesner in \cite{DBLP:conf/lics/CosmoK97}. Kesner and co-authors then explored the connection in various directions \cite{DBLP:journals/mscs/CosmoKP03,DBLP:conf/rta/KesnerL05,DBLP:conf/mfcs/KesnerR09}. In none of these cases, however, do terms and proof nets behave exactly the same.

The graphical representation of $\l$-calculus based on linear logic in \cite{DBLP:conf/csl/AccattoliG09} induced a further calculus with explicit substitutions, the \emph{structural $\l$-calculus} \cite{DBLP:conf/csl/AccattoliK10}, isomorphic to their presentation of proof nets. The structural $\l$-calculus corrects most mentioned mismatches, but it lacks a residual system.

Independently, Milner developed a graphical framework for concurrency, \emph{bigraphs} \cite{DBLP:conf/concur/Milner01}, able to represent the $\pi$-calculus and, consequently, the $\l$-calculus. He extracted from it a calculus with explicit substitutions \cite{DBLP:journals/entcs/Milner07,KesnerOConchuir}, similar in spirit to the structural $\l$-calculus. Accattoli and Kesner later realised that Milner's calculus has a residual system. In 2011-12, they started to work on the LSC, obtained as a merge of Milner's calculus and the structural $\l$-calculus.

At first, the LSC was seen as a minor variation over existing systems. With time, however, a number of properties arose, and the LSC started to be used as a sharp tool for a number of investigations. 
Two of them are relevant for our story. First, the LSC also allows refining the relationship between the $\l$-calculus and the $\pi$-calculus, as shown by the author in \cite{DBLP:journals/corr/abs-1302-6337}. The LSC can then be taken as the harmonious convergence and distillation of three different approaches---linear logic, explicit substitutions, and the $\pi$-calculus---at decomposing the $\l$-calculus. Second, \levy's optimality adapts to the LSC as shown by Barenbaum and Bonelli in \cite{DBLP:conf/rta/BarenbaumB17}, confirming that the advanced rewriting  theory of the $\l$-calculus can indeed be lifted to the micro-step granularity via the LSC.

\subsection{Related Work on Proof Nets} 
The relationship between $\l$-calculi and proof nets has been studied repeatedly, beyond the already cited work (Danos \& Regnier, Kesner \& co-authors, Accattoli \& Guerrini). A nice and detailed introduction to the relationship between $\l$-terms and proof nets is \cite{Gue04netslambda}. 

Laurent extends the translation to represent the $\l\mu$-calculus in \cite{DBLP:journals/tcs/Laurent03,phdlaurent}. In this paper we use an adaptation of his correctness criterion. The translation of differential / resource calculi has also been studied at length: Ehrhard and Regnier \cite{DBLP:journals/entcs/EhrhardR05} study the case without the promotion rule, while Vaux \cite{phdvaux} and  Tranquilli \cite{tranquillithesis,DBLP:journals/tcs/Tranquilli11} include promotion. Vaux also extends the relationship to the classical case (thus encompassing a differential $\l\mu$-calculus), while Tranquilli refines the differential calculus into a \emph{resource calculus} that better matches proof nets. Vaux and Tranquilli use interaction nets to circumvent the minor issue of cuts with axioms. 

Strategies rather than calculi are encoded in interaction nets in \cite{DBLP:conf/ifl/Mackie05}. 

None of these works uses explicit substitutions, so they all suffer of the \emph{mismatch of states}. 
Explicit substitutions are encoded in proof nets in \cite{DBLP:journals/logcom/FernandezS14}, but the operational semantics are not isomorphic, nor correctness is studied. An abstract machine akin to the LSC is mapped to proof nets in \cite{DBLP:conf/csl/MuroyaG17}, but the focus is on cost analyses, rather than on matching syntaxes.

Other works that connect $\l$-calculi and graphical formalisms with some logical background are \cite{DBLP:conf/tlca/AspertiL95,DBLP:conf/lics/GundersenHP13}.

An ancestor of this paper is \cite{DBLP:journals/tcs/Accattoli15}, that adopts essentially the same syntax for proof nets. In that work, however, the operational semantics is small-step rather than micro-step, there is no study of the quotient, and no use of contexts, nor it deals with the LSC.
\section{The Linear Substitution Calculus}
\paragraph{Expressions and terms.} One of the features of the LSC is the use of contexts to define the rewriting rules. Contexts are terms with a single occurrence of a special constructor called \emph{hole}, and often noted $\ctxhole$, that is a placeholder for a removed subterm. To study the relationship with proof nets, it is necessary to represent both terms and contexts, and, to reduce the number of cases in definitions and proofs,  we consider a syntactic category generalizing both. \emph{Expressions} may have 0, 1, or more holes. Proof nets also require holes to carry the set $\nameset$ of variables that can appear free in any subterm replacing the hole---\eg $\nameset = \set{ \var, \vartwo, \varthree}$. Expressions are then defined as follows:
\begin{center}$\begin{array}{r@{\hspace{1cm}}rcl}
 \textsc{Expressions} & \expr,\exprtwo,\exprthree,\exprfour & \grameq & \var\mid  \ctxholens \mid \la\var\expr \mid \expr\exprtwo \mid \expr\esub\var\exprtwo 
\end{array}$\end{center}
\emph{Terms} are  expressions without holes, noted $\tm$, $\tmtwo$, $\tmthree$, and so on, and \emph{contexts} are expressions with exactly one hole, noted $\ctx$, $\ctxtwo$, $\ctxthree$, etc. 

The construct $ \tm\esub\var\tmtwo$ is an \emph{explicit substitution}, shortened \emph{ES}, of $\tmtwo$ for $\var$ in $\tm$---essentially, it is a more compact notation for $\letexp\var\tmtwo\tm$. Both $\la\var\tm$ and $\tm\esub\var\tmtwo$ bind $\var$ in $\tm$. Meta-level, capture-avoiding substitution is rather noted $\tm\isub\var\tmtwo$. On terms, we silently work modulo $\alpha$-equivalence, so that for instance $(\la\var ((\var\vartwo\varthree) \esub\vartwo\var)\isub\varthree{\var\vartwo} = \la{\var'} ((\var'\vartwo'(\var\vartwo)) \esub{\vartwo'}{\var'})$. Applications associate to the left. Free variables of holes are defined by $\fv\ctxholens \defeq \nameset$, and for the other constructors as expected. The \emph{multiplicity} of a variable $\var$ in a \emph{term} $\tm$, noted $\sizep\tm\var$, is the number of free occurrences of $\var$ in $\tm$. 

\paragraph{Contexts.} The LSC uses contexts extensively, in particular \emph{substitution contexts}: 
\begin{center}
$\begin{array}{r@{\hspace{.5cm}}rcl}
 \textsc{Substitution contexts} & \sctx, \sctxtwo,\sctxthree 	& \grameq & \ctxholens \mid \sctx\esub\var\tm 
\end{array}
$\end{center}
Sometimes we write $\ctxns$ for a context $\ctx$ whose hole $\ctxholens$ is annotated with $\nameset$, and we call $\nameset$ the \emph{interface} of $\ctx$. Note that the free variables of $\ctxns$ do not necessarily include those in its interface $\nameset$, because the variables in $\nameset$ can be captured by the binders in $\ctxns$. 

The basic operation over contexts is \emph{plugging} of an expression $\expr$ in the hole of the context $\ctx$, that produces the expression $\ctxp\expr$. The operation is defined only when the free variables $\fv\expr$ of $\expr$ are included in the interface of the context. 
\begin{center}
$\begin{array}{rcl@{\hspace{1.5cm}}rcl}
\multicolumn{6}{c}{\textsc{Plugging of $\expr$ in $\ctxns$ (assuming $\fv\expr\subseteq \nameset$)}}
\\[.2cm]
\ctxholensp \expr & \defeq & \expr 
&
(\la\var\ctx)\ctxholep\expr & \defeq &  \la\var\ctxp\expr
\\
(\ctx \tmtwo) \ctxholep\expr & \defeq &  \ctx \ctxholep\expr \tmtwo
&
(\tmtwo \ctx) \ctxholep\expr & \defeq &  \tmtwo \ctx \ctxholep\expr
\\
(\ctx \esub\var\tmtwo) \ctxholep\expr & \defeq &  \ctxp\expr \esub\var\tmtwo
&
(\tmtwo \esub\var\ctx) \ctxholep\expr & \defeq &  \tmtwo \esub\var{\ctxp\expr}
\end{array}
$\end{center}
An example of context is $\ctxn{\set{\var,\vartwo}} \defeq \la\var (\vartwo \ctxholen{\set{\var,\vartwo}}\esub\varthree\var)$, and one of plugging is $\ctxn{\set{\var,\vartwo}}\ctxholep{\var\var} = \la\var (\vartwo (\var\var)\esub\varthree\var)$. Note the absence of side conditions in the cases for $\la\var\ctx$ and $\ctx \esub\var\tmtwo$---it means that plugging in a context can capture variables, as in the given example.
Clearly, $\ctxp\expr$ is a term / context if and only if $\expr$ is a term / context. Note also that if $\tm$ is a term and $\tmtwo$ is a subterm of $\tm$ then $\tm = \ctxp\tmtwo$ for some context $\ctx$. Such a context $\ctx$ is unique up to the annotation $\nameset$ of the hole of $\ctx$, which only has to satisfy $\fv\tmtwo \subseteq \nameset$, and that can always be satisfied by some $\nameset$.

We also define \emph{the set $\cv \ctxns$ of variables captured by a context $\ctxns$}:
\begin{center}
$\begin{array}{r\colspace c\colspace l}
\multicolumn{3}{c}{\textsc{Variables captured by a context}}
\\[.2cm]
\cv\ctxholens & \defeq & \emptyset
\\
\cv {\la\var\ctxns}  = \cv {\ctxns\esub\var\tm} & \defeq & \cv\ctxns \cup \set\var

\\
\cv{\tm\ctxns} = \cv{\ctxns\tm} = \cv{\tm\esub\var\ctxns} & \defeq &  \cv\ctxns
\end{array}
$\end{center}

\paragraph{Rewriting rules for terms.} The rewriting rules of the LSC concern terms only. They are unusual as they use contexts in two ways: to allow their application anywhere in the term---and this is standard---and to define the rules at top level---this is less common (note the substitution context $\sctx$ and the context $\ctx$ in rules $\tom$ and $\toe$ below). We write $\ctxfp\tm$ if $\ctx$ does not capture any free variable of $\tm$, that is, if $\cv\ctx \cap \fv\tm = \emptyset$.
\begin{center}
 \textsc{Rewriting rules}\medskip
 
$\begin{array}{r@{\hspace{1cm}}rclllllll}
    \textsc{Multiplicative} &\sctxp{\la\var\tm} \tmtwo & \todb &  \sctxp {\tm \esub\var\tmtwo }
    \\
    \textsc{Milner exponential} & \ctxfp\var \esub\var\tmtwo & \tols & \ctxfp\tmtwo \esub\var\tmtwo &
    \\
    \textsc{Garbage collection} & \tm \esub\var\tmtwo & \togc & \tm & \mbox{if $\var \notin \fv\tm$}
    \\[.2cm]
    \textsc{Contextual closures} & 	
	\AxiomC{$\tm \Rew{a} \tm'$}		
	\UnaryInfC{$\ctxp\tm \Rew{a} \ctxp{\tm'}$}
	\DisplayProof   
    && \mbox{for $a \in \set{\msym,\esym,\gcsym}$}    
    \\[.3cm]
    \textsc{Notation} & \tolsc  & \defeq & \todb \cup \tols \cup \togc
\end{array}$
\end{center}
Note that in $\tom$ (resp. $\toe$) we assume that $\sctx$ (resp. $\ctx$) does not capture variables in $\fv\tmtwo$---this is always possible by a (on-the-fly) $\alpha$-renaming of $\sctxp{\la\var\tm}$ (resp. $\ctxfp\var$), as we work modulo $\alpha$. Similarly the interface of $\ctx$ can always be assumed to contain $\fv\tmtwo$.

\paragraph{Structural equivalence.} The LSC is sometimes enriched with the following notion of structural equivalence $\eqstruct$ \cite{DBLP:conf/popl/AccattoliBKL14}.

\begin{definition}[Structural equivalence]
Structural equivalence $\eqstruct$ is defined as the symmetric, reflexive, transitive, and contextual closure of
the following axioms:
\begin{center}
$\begin{array}{rll@{\hspace{1em}}l}
    (\la\vartwo\tm)\esub{\var}{\tmtwo}             & \eqstruct_{\l} & \la\vartwo\tm\esub{\var}{\tmtwo}               & \text{if $\vartwo \not\in \fv{\tmtwo}$} \\
    (\tm\,\tmthree)\esub{\var}{\tmtwo}             & \eqstruct_{@l} & \tm\esub{\var}{\tmtwo}\,\tmthree               & \text{if $\var \not\in \fv{\tmthree}$} \\
    \tm\esub{\var}{\tmtwo}\esub{\vartwo}{\tmthree} & \eqstruct_{com} & \tm\esub{\vartwo}{\tmthree}\esub{\var}{\tmtwo} & \text{if $\vartwo \not\in \fv{\tmtwo}$ and $\var \not\in \fv{\tmthree}$}
\end{array}$
\end{center}
\end{definition}

Its key property is that it commutes with evaluation in the following strong sense.

\begin{proposition}[$\eqstruct$ is a strong bisimulation wrt $\tolsc$ \cite{DBLP:conf/popl/AccattoliBKL14}]
	\label{prop:bisimulation}
	Let $a\in\set{\msym,\esym,\gc}$. If $\tm \eqstruct\tmtwo \Rew{a} \tmthree$ then exists $\tmfour$ such that $\tm \Rew{a}\tmfour \eqstruct\tmthree$.
\end{proposition}

Essentially, $\equiv$ never creates redexes, it can be postponed, and vanishes on normal forms (that have no ES). We are going to prove that $\equiv$ is exactly the quotient induced by translation to proof nets (\refthm{quotient}, page \pageref{thm:quotient}). The absence of the axiom $(\tm\,\tmthree)\esub{\var}{\tmtwo} \eqstruct_{@r} \tm\,\tmthree\esub{\var}{\tmtwo}$ if $\var \not\in \fv{\tm}$ is correct: the two terms do not have the same proof net representation (defined in the next section), moreover adding this axiom to $\equiv$ breaks \refprop{bisimulation}. The extension with $\eqstruct_{@r}$ has nonetheless been studied in \cite{DBLP:journals/corr/abs-1203-0670}.

\section{Proof Nets}
\label{sect:proofnets}
\begin{figure}[t]
\centering
\input{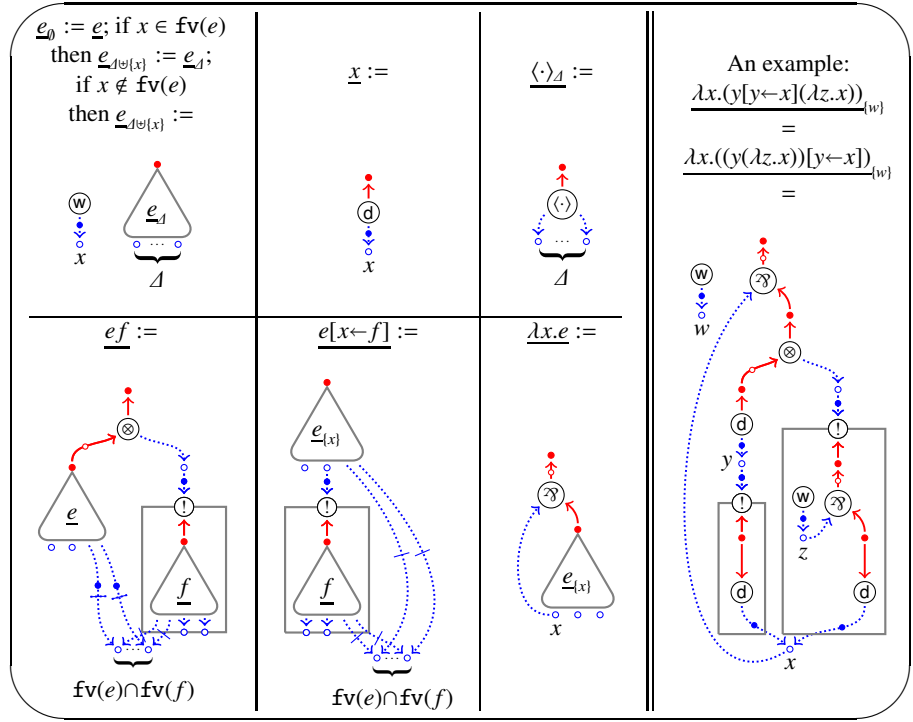}
\caption{\label{fig:trans} Translation of expressions to nets, plus an example of translation.}
\end{figure}

\paragraph{Introduction.} Our presentation of proof nets, similar to the one in \cite{DBLP:journals/tcs/Accattoli15}, is nonstandard in at least four points---we suggest to have a quick look to \reffig{trans}, page \pageref{fig:trans}:
\begin{enumerate}
\item \emph{Hyper-graphs}: we use directed hyper-graphs (for which formulas are nodes and links---\ie\  logical rules---are hyper-edges) rather than the usual graphs with pending edges (for which formulas are edges and links are nodes). We prefer hyper-graphs---that despite the scaring name are nothing but bipartite graphs---because they give
\begin{enumerate}
\item \emph{Contraction algebra for free}: contraction is represented modulo commutativity, associativity, and permutation with box borders \emph{for free}, by admitting that exponential nodes can have more than one incoming link,
\item \emph{Cut-axiom quotient for free}: cut and axiom links are represented implicitly, collapsing them on nodes. This is analogous to what happens in interaction nets. Intuitively, our multiplicative nodes are \emph{wires}, with exponential nodes being \emph{hyper}-wires, \ie wires involving an arbitrary number of ports;
\item \emph{Subnets as subsets}: subnets can be elegantly defined as subsets of links, which would not be possible when adopting other approaches such as generalized $?$-links or a standard interaction nets formalism without hyper-wires.
\end{enumerate}
The choice of hyper-graphs, however, has various (minor) technical consequences, and the formulation of some usual notions (\eg\ the nesting condition for boxes) shall be slightly different with respect to the literature.

\item \emph{Directed links and polarity}: our links are directed and we apply a correctness criterion based on directed paths. Be careful, however, that we do not follow the usual premises-to-conclusions orientation for links, nor the input-output orientation sometimes at work for $\l$-calculi or intuitionistic settings. We follow, instead, the orientation induced by logical polarity according to Laurent's correctness criterion for polarised proof nets \cite{DBLP:journals/tcs/Laurent03,phdlaurent}. Let us point out that Laurent defines proof nets using the premises-to-conclusions orientation and then he switches to the polarised orientation for the correctness criterion. We prefer to adopt only one orientation, the polarised one, which we also employ to define proof nets.

\item \emph{Syntax tree}: since we use proof nets to represent terms, we arrange them on the plane according to the syntax tree of the corresponding terms, and not according to the corresponding sequent calculus proof, analogously to the graph rewriting literature on the $\l$-calculus (\eg \cite{Wad:SemPra:71}) but in contrast to the linear logic literature.

\item \emph{Contexts}: to mimic the use of contexts in the LSC rewriting rules, we need to have a notion of context net. Therefore, we have a special link for context holes.

\end{enumerate}

\begin{figure}[t]
\centering

\renewcommand{\stalt}{30pt}
\renewcommand{\stlar}{26pt}

\ovalbox{
\centering
\begin{tabular}{c}
\begin{tabular}{\spc c\spc\spc c\spc\spc c\spc\spc c\spc\spc c\spc}
\small Weakening & \small Dereliction & \small Bang & \small Tensor & \small Par
\\[.1cm]
\begin{tikzpicture}[ocenter]
\node at (0,0) [eportn, label=below:$\esym$](up){};

\lweakn{up}{weak}
\end{tikzpicture}
&
\begin{tikzpicture}[ocenter]
\node at (0,0) [label=above:$\msym$, mportn](up){};
\node at (up.center) [below=\stalt, eportn,label=below:${?\msym^\bot=\esym}$](down){};
\ldern{down}{up}{der}
\end{tikzpicture}
&
\begin{tikzpicture}[ocenter]
\node at (0,0) [label=above:${?\msym^\bot=\esym}$, eportn](up){};
\node at (up.center) [below=\stalt, label=below:$\msym$, mportn](down){};
\lbangn{down}{up}{bang}
\end{tikzpicture}
&
\begin{tikzpicture}[ocenter]
\node at (0,0) [label=above:$\msym$, mportn](up){};
\node at (up.center) [below right=\stalt and \hstlar,label=below:$\esym$,  eportn](downr){};
\node at (up.center) [below left=\stalt and \hstlar,  label=below:${\msym=\esym\parr \msym}$, mportn](downl){};
\ltensn{up}{downl}{downr}{tens}
\end{tikzpicture}
&

\begin{tikzpicture}[ocenter]
\node at (0,0) [ label=above:${\msym=\esym \parr \msym}$, mportn](up){};
\node at (up.center) [below right=\stalt and \hstlar, label=below:$\msym$, mportn](downr){};
\node at (up.center) [below left=\stalt and \hstlar, label=below:$\esym$, eportn](downl){};
\lparn{up}{downl}{downr}{par}
\end{tikzpicture}
\end{tabular}

\\[1.1cm]
\hline
\\[-.2cm]

\begin{tabular}{c\colspace c\colspace c\colspace c\colspace c\colspace |c\colspace c\colspace ccccc}
\small Context hole &\small Collapsed box
\\[.1cm]

\begin{tikzpicture}[ocenter]
\node at (0,0) [label=above:$\msym$, mportn](up){};
\node at (up.center) [below right=\stalt and \hstlar, label=below:$\esym$, eportn](downr){};
\node at (up.center) [below left=\stalt and \hstlar, label=below:$\esym$, eportn](downl){};
\gdots{downl}{downr}{}
\lcontextn{up}{downl}{downr}{par}
\end{tikzpicture}

&
\begin{tikzpicture}[ocenter]
\node at (0,0) [label=above:$\esym$, eportn](up){};
\node at (up.center) [below right=\stalt and \hstlar, label=below:$\esym$, eportn](downr){};
\node at (up.center) [below left=\stalt and \hstlar, label=below:$\esym$, eportn](downl){};
\gdots{downl}{downr}{}
\lcolboxn{up}{downl}{downr}{par}
\end{tikzpicture}

\end{tabular}
\end{tabular}
}
\renewcommand{\stalt}{22pt}
\renewcommand{\stlar}{15pt}
\caption{\label{fig:links} Links.}
\end{figure}

\paragraph{Nets.} We first overview some choices and terminology.
\begin{itemize}
  \item \emph{Hyper-graphs}: nets are directed and labelled hyper-graphs $G=(\nodeset G, \linkset G)$, \ie, graphs where $\nodeset G$ is a set of labelled \emph{nodes} and $\linkset G$ is a set of labelled and \emph{directed hyper-edges}, called \emph{links}, which are edges with 0, 1, or more sources and 0, 1, or more targets\footnote{
A hyper-graph $G$ can be understood as a bipartite graph $B_G$, where $V_1(B_G)$ is $\nodeset G$ and $V_2(B_G)$ is $ \linkset G$, and the edges are determined by the relations \textit{being a source} and \textit{being a target} of a hyper-edge.}.

\item \emph{Nodes}: nodes are labelled with a type in $\set{\esym,\msym}$, where $\esym$ stands for \textit{exponential} and $\msym$ for \textit{multiplicative}. If a node $\nnode$ has type $\esym$ (resp. $\msym$) we say that it is a $\esym$-node (resp. $\msym$-node). The label of a node is usually left implicit, as $\esym$ and $\msym$ nodes are distinguished graphically, using both colours and different shapes: $\esym$-nodes are cyan and white-filled, while $\msym$-nodes are brown and dot-like. We come back to types below.

\item \emph{Links}: we consider hyper-graphs whose links are labelled from $\set{\bang,\der,\weak,\parr,\otimes, \ctxhole, \genax}$, corresponding to the promotion, dereliction, weakening, par, and tensor rules of linear logic, plus a link $\ctxhole$ for context holes and a link $\genax$ used for defining the correction graph---contraction is hard-coded on nodes, as already explained. The label of a link $\link$ forces the number and the type of the source and target nodes of $\link$, as shown in \reffig{links} (types shall be discussed next). Similarly to nodes, we use colours and shapes for the type of the source/target connection of a link to a node: $\esym$-connections are blue and dotted, while $\msym$-connections are red and solid. Our choice of shapes allows reading the paper also if printed in black and white. 

\item \emph{Principal conclusions}: note that every link except $\ctxhole$ and $\genax$ has exactly one connection with a little circle: it denotes the \emph{principal} node, \ie\ the node on which the link can interact. Notice the principal node for tensor and $\oc$, which is not misplaced. 

\item \emph{Typing}: nets are typed using a recursive type, usually noted $o=\oc o\multimap o$, but that we rename $\msym=\oc\msym\multimap \msym=?\msym^\bot\parr \msym$ because $\msym$ is a mnemonic for \emph{multiplicative}. Let $\esym \defeq ?\msym^\bot$, where $\esym$ stands for  \emph{exponential}. Note that $\msym=\esym^\bot\multimap \msym = \esym\parr \msym$. Links are typed using $\msym$ and $\esym$, but the types are omitted by all figures except \reffig{links} because they are represented using colours and with different shapes ($\msym$-nodes are brown and dot-like, $\esym$-nodes are white-filled cyan circles). Let us explain the types in \reffig{links}. They may be counter-intuitive at first: note in particular the $\oc$ and $\tens$ links, that have an unexpected type on their logical conclusion---it simply has to be negated, because the expected orientation would be the opposite one.

\item \emph{More on nodes}: a node is \emph{initial} if it is not the target of any link; \emph{terminal} if it is not the source of any link; \emph{isolated} if it is initial and terminal; \emph{internal} if it is not initial nor terminal.

\item \emph{Boxes}: every $\bang$-link has an associated \emph{box}, \ie, a sub-hyper-graph of $\net$ (have a look at \reffig{trans}), meant to be a sub-net.
\item \emph{Context holes and collapsed boxes}: it is natural to wonder if $\ctxhole$ and $\genax$ links can be merged into a single kind of link. They indeed play very similar roles, except that they have different polarised typings, which is why we distinguish them. 
\end{itemize}

We first introduce \emph{pre-nets}, and then add boxes on top of them, obtaining \emph{nets}:

\begin{definition}[Pre-nets]
\label{def:nets}
	A \emph{pre-net} $\net$ is a triple $(|\net|, \fv{\net}, \rootnode_\net)$, where $|\net|$ is a hyper-graph $(\nodeset\net,\linkset\net)$ whose nodes are labelled with either $\esym$ or $\msym$ and whose hyper-edges are $\set{\oc,\der,\weak,\parr,\otimes, \ctxhole, \genax}$-links, and such that:
	\begin{itemize}
		\item \emph{Root}: $\rootnode_\net\in \nodeset \net$ is a terminal $\msym$-node of $\net$, called the \emph{root} of $\net$.
		\item \emph{Free variables}: $\fv{\net}$ is the set of terminal $\esym$-nodes of $\net$, also called \emph{free variables} of $\net$, which are targets of $\set{\der,\weak,\ctxhole,\genax}$-links (\ie\ they are not allowed to be targets of $\otimes$-links, nor to be isolated).

	\item \emph{Nodes}: every node has \emph{at least} one incoming link and \emph{at most one} outgoing link. Moreover,
	\begin{itemize}
		\item \emph{Multiplicative}: $\msym$-nodes have \emph{exactly one} incoming link;
		\item \emph{Exponential}: if an $\esym$-node has more than one incoming link then they are $\der$-links.
	\end{itemize}
	\end{itemize}
\end{definition}

\begin{definition}[Nets]
	A \emph{net} $\net$ is a pre-net together with a function $\bbox_\net$ (or simply $\bbox$) associating to every $\oc$-link $\link$ a subset $\bboxp{\link}$ of $\linkset \net \setminus\set{\link}$ (\ie\ the links of $\net$ except $\link$ itself), called the \emph{interior of the box} of $\link$, such that $\bboxp{\link}$ is a pre-net verifying (explanations follow):	
		\begin{itemize}
			\item \emph{Border}: the root $\rootnode_{\bboxp{\link}}$ is the source $\msym$-nodes of $\link$, and any free variable of $\bboxp{\link}$ is not the target of a weakening.
			\item \emph{Nesting}: for any $\oc$-box $\bboxp{\linktwo}$ if $\bboxp{\link}$ and $\bboxp{\linktwo}$ have non-empty intersection---that is, if $\emptyset\neq I\defeq|\bboxp{\link}|\cap|\bboxp{\linktwo}|$---and one is not entirely contained in the other---that is, if $|\bboxp{\link}|\not \subseteq|\bboxp{\linktwo}|$, and $|\bboxp{\linktwo}|\not \subseteq|\bboxp{\link}|$---then all the nodes in $I$ are free variables of both $\bboxp{\link}$ and $\bboxp{\linktwo}$.
			\item \emph{Internal closure}: 
			\begin{itemize}
			  \item \emph{Contractions}: if a contraction node is internal to $\bboxp{\link}$ then all its premises are in $\bboxp{\link}$---formally, $\linktwo\in\bboxp{\link}$ for any link $\linktwo$ of $\net$ having as target an internal $\esym$-node of $\bboxp{\link}$.

			  \item \emph{Boxes}: $\bboxp\linktwo\subseteq\bboxp\link$ for any $\oc$-link $\linktwo \in\bboxp\link$.
			\end{itemize}
		\end{itemize}
	A net is 
	\begin{itemize}
	\item a \emph{term net} if it has no $\set{\ctxhole,\genax}$-links; 
	\item a \emph{context net} if it has exactly one $\ctxhole$-link; 
	\item a \emph{correction net} if it has no $\oc$-links. 
	\end{itemize}
	As for the calculus, the \emph{interface} of a $\ctxhole$-link is the set of its free variables, and the interface of a context net is the interface of its $\ctxhole$-link.
\end{definition}

\begin{remark}
Comments on the definition of net:
\begin{enumerate}
\item \emph{Weakenings and box borders}: in the border condition for nets the fact that the free variables are not the target of a weakening means that weakenings are assumed to be pushed out of boxes as much as possible---of course the rewriting rules shall have to preserve this invariant.

\item \emph{Weakenings are not represented as nullary contractions}: given the representation of contractions, it would be tempting to define weakenings as nullary contractions. However, such a choice would be problematic with respect to correctness (to be defined soon), as it would introduce many initial $\esym$-nodes in a correct net and thus blur the distinction between the root of the net, supposed to represent the output and to be unique (in a correct net), and substitutions on a variable with no occurrences (\ie\ weakened subterms), that need not to be unique.

\item \emph{Internal closure wrt contractions}: it is a by-product of collapsing contractions on nodes, which is also the reason for the unusual formulation of the nesting condition. In fact, two boxes that are intuitively disjoint can in our syntax share free variables, because of an implicit contraction merging two of them, as in the example in \reffig{trans}.

\item \emph{Boxes as nets}: note that a box $\bboxp\link$ in a net $\net$ is only a \emph{pre}-net, by definition. Every box in a net $\net$, however, inherits a net structure from $\net$. Indeed, one can restrict the box function $\bbox_\net$ of $\net$ to the $\oc$-links of $\bboxp\link$, and see $\bboxp\link$ as a \emph{net}, because all the required conditions are automatically satisfied by the internal boxes closure and by the fact that such boxes are boxes in $\net$. Therefore, we freely consider boxes as \emph{nets}.

\item \label{def:arguments}
\emph{Tensors and $\oc$-boxes}: the requirements that the $\esym$-target of a $\tens$-link cannot be the free variable of a net, nor the target of more than one link force these nodes to be sources of $\oc$-links. Therefore, every $\tens$-link is paired to a $\oc$-link, and thus a box.

\item \emph{Acyclic nesting}: the fact that a $\oc$-link does not belong to its box, plus the internal closure condition, imply that the nesting relation between boxes cannot be cyclic, as we now show. Let $\link$ and $\linktwo$ be $\oc$-links. If $\link \in \bboxp{\linktwo}$ then by internal closure $\bboxp{\link} \subseteq \bboxp{\linktwo}$. It cannot then be that $\linktwo \in \bboxp{\link}$, otherwise $\link$ would belong to its own box, because $\link \in \bboxp{\linktwo} \subseteq \bboxp{\link}$ by internal closure. 
\end{enumerate}
\end{remark}

\paragraph{Terminology about nets.} Some further terminology and conventions:
\begin{itemize}
  \item \label{def:level}
  The \emph{level} of a node/link/box is the maximum number of nested boxes in which it is contained\footnote{Here the words \emph{maximum} and \emph{nested} are due to the fact that the free variables of $\oc$-boxes may belong to two not nested boxes, as in the example in \reffig{trans}, because of the way we represent contraction.} (a $\oc$-link is not contained in its own box). Note that the level is well defined by the acyclicity of nesting just pointed out. In particular, if a net has $\oc$-links then it has at least one $\oc$-link at level 0.
  
  \item A \emph{variable} $\var$ is a $\esym$-node that is the target of a $\set{\der,\weak}$-link---equivalently, that is not the target of a $\tens$-link.
  
  \item Two links are \emph{contracted} if they share an $\esym$-target. Note that the exponential condition states that only derelictions (\ie\ $\der$-links) can be contracted. In particular, no link can be contracted with a weakening. 
  
  \item A \emph{free weakening} in a net $\net$ is a weakening whose node is a free variable of $\net$. 
  
  \item The \emph{multiplicity} of a variable $\var$ in $\net$, noted $\sizep\net\var$, is 0 if $\var$ is the target of a weakening, and $n\geq 1$ if it is the target of $n$ derelictions.
  
  \item Sometimes (\eg\ the bottom half of \reffig{trans}), the figures show a link in a box having as target a contracted $\esym$-node $x$ which is outside the box: in those cases $x$ is part of the box, it is outside of the box only in order to simplify the representation.
\end{itemize}

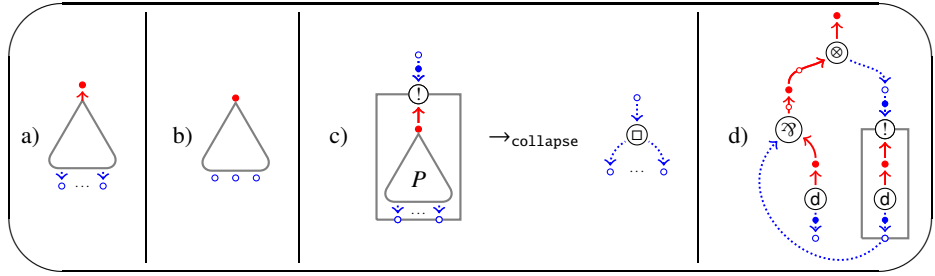
\begin{figure}[t]
\centering
\ovalbox{\begin{tabular}{c\spc|\spc c\spc |\spc c\spc |\spc c}
a)
\begin{tikzpicture}[ocenter]
\node at (0,0) [mportn](up){};
\inetcell[inductiveTr,at=(up.center), below=\sepbox](body){}[180]    
\node at (body.left pax) [eportn, below=\sepbox](freevarleft){};
\node at (body.right pax) [eportn, below=\sepbox](freevarright){};

\draw[exppuren] (body.left pax) to (freevarleft);
\draw[exppuren] (body.right pax) to (freevarright);
\draw[puren] (body.pal) to (up) ;
\gdots{freevarleft}{freevarright}{}
\end{tikzpicture}
&
b)
\begin{tikzpicture}[ocenter]
\node at (0,0) [mportn](up){};
\inetcell[inductiveTr,at=(up.center), below=\sepboxshort](body){}[180]    
\node at (body.left pax) [eportn, below=\sepboxshort](freevarleft){};
\node at (body.middle pax) [eportn, below=\sepboxshort](freevarmiddle){};
\node at (body.right pax) [eportn, below=\sepboxshort](freevarright){};
\end{tikzpicture}
&

\begin{tabular}{c\spc c\spc c\spc c}
c)&
\begin{tikzpicture}[ocenter]
\node at (0,0) [eportn](up){};
\node at (up.center) [below =\stalt, mportn](downr){};

\lbangn{downr}{up}{bang}
\inetcell[inductiveTr,at=(downr.center), below=\sepboxshort](body){\footnotesize$\net$}[180]    
\node at (body.left pax) [eportn, below=\sepbox](x){};
\node at (body.right pax) [eportn, below=\sepbox](freevarright){};

\aBoxWithPalOnTopAndVeticalNodeBound{bang}{freevarright}{exbox}{15pt}{15pt}
\lbangn{downr}{up}{bang}
\node at (body.left pax) [eportn, below=\sepbox](freevarleft){};

\node at (body.right pax) [eportn, below=\sepbox](freevarright){};
\draw[exppuren] (body.left pax) to (freevarleft);
\draw[exppuren] (body.right pax) to (freevarright);

\gdots{body.left pax}{body.right pax}{below=.5pt}
\end{tikzpicture}
&
$\Rew{\tt collapse}$
&
\begin{tikzpicture}[ocenter]
\node at (0,0) [eportn](up){};
\node at (up.center) [below right=\stalt and \hstlar, eportn](downr){};
\node at (up.center) [below left=\stalt and \hstlar, eportn](downl){};
\gdots{downl}{downr}{}
\lcolboxn{up}{downl}{downr}{par}
\end{tikzpicture}
\end{tabular}
&
\begin{tabular}{c\spc c}
d)
&
\begin{tikzpicture}[ocenter]

\node at (0,0) [mportn](tens_output){};
\node at (tens_output.center) [below right=\stalt and 1.8*\hstlar, eportn](right_bang_output){};
\node at (tens_output.center) [below left=\stalt and 1.8*\hstlar, mportn](head_lam_output){};
\ltensn{tens_output}{head_lam_output}{right_bang_output}{ap};

\node at (right_bang_output.center) [below=\stalt, mportn](boxed_var_output){};
\lbangn{boxed_var_output}{right_bang_output}{right_bang};

\node at (boxed_var_output.center) [below=\stalt, eportn](boxed_var_input){};
\ldern{boxed_var_input}{boxed_var_output}{arg_var}

\node at (boxed_var_input.center) [above=5pt, nospace](right_box_bottom_point){};
\aBoxWithPalOnTopAndVeticalNodeBound{right_bang}{boxed_var_input}{exbox}{8pt}{8pt}
\lbangn{boxed_var_output}{right_bang_output}{right_bang};
\node at (boxed_var_output.center) [below=\stalt, eportn](boxed_var_input){};

\node at (head_lam_output.center) [below right=\stalt and \hstlar, mportn](head_var_output){};

\node at (head_var_output.center) [below =\stalt, eportn](head_var_input){};
\ldern{head_var_input}{head_var_output}{head_var}

\lparnangle{head_lam_output}{boxed_var_input}{head_var_output}{head_lam}{out=-135, in=-135, looseness=1.5, overlay}

\node at (head_var_input.center) [below =8pt, nospace](ghost_node){};
\end{tikzpicture}
\end{tabular}

\end{tabular}
}
\caption{\label{fig:various} Various images.}
\end{figure}

\paragraph{Translation.} Nets representing terms have the general form in \reffig{various}.a, also represented as in \reffig{various}.b. The translation $\lamtonets{\cdot }$ from expression to nets is in \reffig{trans}. 

A net which is the translation of an expression is a \emph{proof net}. Note the example in \reffig{trans}: two different terms translate to the same proof net, showing that proof nets quotient LSC terms.

The translation $\lamtonets{\cdot}$ is refined to a translation $\lamtonetsvar{\cdot }{\varset}$, where $\varset$ is a set of variables, in order to properly handle weakenings during cut-elimination. The reason is that an erasing step on terms simply erases a subterm, while on nets it also introduces some weakenings: without the refinement the translation would not be stable by reduction. 

Note that in some cases there are various edges entering an $\esym$-node, that is the way we represent contraction. In some cases the $\esym$-nodes have an incoming connection with a perpendicular little bar: it represents an arbitrary number ($>0$) of incoming connections. Structurally equivalent terms are translated to the same proof net, see \reffig{quotient} at page \pageref{fig:quotient}.

\paragraph{$\alpha$-Equivalence.} To circumvent an explicit and formal treatment of $\alpha$-equivalence we assume that the set of $\esym$-nodes and the set of variable names for terms coincide. This convention removes the need to label the free variables of $\lamtonetsvar{\tm}{\varset}$ with the name of the corresponding free variables in $\tm$ or $\varset$. Actually, before translating a term $\tm$ it is necessary to pick a \emph{well-named} $\alpha$-equivalent term $\tm'$, \ie\ a term such that any two different variables (bound or free) have different names.

\paragraph{Paths.} A \emph{path} $\path$ of length $k\in\nat$ from $\nnode$ to $\nnodetwo$, noted $\path:\nnode\to^k \nnodetwo$, is an alternated sequence of nodes and links $\nnode = \nnode_1,\link_1,\ldots,\link_k, \nnode_{k+1} = \nnodetwo$ such that link $\link_i$ has source $\nnode_i$ and target $\nnode_{i+1}$ for $i\in\set{1,\ldots,k}$. A \emph{cycle} is a path $\nnode\to^k \nnode$ with $k>0$.

\paragraph{Correctness.} The correctness criterion is an adaptation of Laurent's criterion for polarized nets, and it is the simplest known criterion for proof nets. It is based on the notion of correction net, which---as usual for nets with boxes---is obtained by collapsing boxes into generalized axiom links, \ie\ our $\genax$-links (see \reffig{links}).

\begin{definition}[Correction net]
Let $\net$ be a net. The \emph{correction net} $\zeronet{\net}$ of $\net$ is the net obtained from $\net$ by collapsing each $\oc$-box at level 0 in $\net$ into a $\genax$-link with the same interface, by applying the rule in \reffig{various}.c. 
\end{definition}


\begin{definition}[Correctness]
  \label{def:correctness}
	A net $\net$ is \emph{correct} if:
	\begin{itemize}
		\item \emph{Root}: the root of $\net$ induces the only terminal $\msym$-node of $\zeronet{\net}$.
		\item \emph{Acyclicity}: $\zeronet{\net}$ is acyclic.
		\item \emph{Recursive correctness}: the box of every $\oc$-link at level 0 is correct.
	\end{itemize}
\end{definition}

An example of net that is not correct is in \reffig{various}.d: the correction net obtained by collapsing the box indeed has a cycle.

Note that acyclicity provides an induction principle on correct nets, because it implies that there is a maximal length for paths in the correction net associated to the net. 

\paragraph{Proof nets are correct.} As usual, an easy and omitted induction on the translation shows that the translation of an expression is correct, \ie\ that:

\begin{proposition}[Proof nets are correct]
\label{prop:transl-is-correct} 
Let $\expr$ be an expression and $\nameset$ a set of variables. Then $\lamtonetsvar{\expr}{\varset}$ is a correct net of free variables $\fv\expr \cup \varset$. Moreover, 
\begin{enumerate}
  \item \label{p:transl-is-correct-terms}
  if $\expr$ is a term then $\lamtonetsvar{\expr}{\varset}$ is a term net and their variables have the same multiplicity, that is, $\sizep\expr\var = \sizep{\lamtonetsvar{\expr}{\varset}}\var$ for every variable $\var$.
  
  \item \label{p:transl-is-correct-contexts}
  if $\expr$ is a context then $\lamtonetsvar{\expr}{\varset}$ is a context net.
\end{enumerate}
\end{proposition}


\paragraph{Linear skeleton.} We have the following strong structural property.

\begin{toappendix}
\begin{lemma}[Linear skeleton]
\label{l:linear-subnet}
Let $\net$ be a correct net. The \emph{linear skeleton} of $\zeronet{\net}$, given by $\msym$-nodes and the red (or linear) paths between them, is a linear order.
\end{lemma}
\end{toappendix}

%
%

\section{Sequentialisation and Quotient}
\label{sect:sequentialisation}
In this section we prove the sequentialisation theorem and the fact that the quotient induced by the translation on terms is exactly the structural equivalence $\eqstruct$ of the LSC.

\paragraph{Subnets.} The first concept that we need is the one of \emph{subnet} $\nettwo$ of a correct net $\net$, that is a subset of the links of $\net$ plus some closure conditions. These conditions avoid that $\nettwo$ prunes the interior of a box in $\net$, or takes part of the interior without taking the whole box, or takes only some of the premises of an internal contraction.

For the sake of simplicity, in the following we specify sub-hyper-graphs of a net by simply specifying their set of links. This is an innocent abuse, because---by definition of (pre-)net---there cannot be isolated nodes, and so the set of nodes is retrievable from the set of links. Similarly, the boxes of $\oc$-links are inherited from the net.

\begin{definition}[Subnet]
	Let $\net$ be a correct net. A \emph{subnet} $\nettwo$ of $\net$ is a subset of its links such that it is a correct net (with respect to the $\bbox$ function inherited from $\net$) and satisfies the following closure conditions:
	\begin{itemize}
		\item \emph{Contractions}: $\link\in\nettwo$ for any link $\link$ of $\net$ having as target an internal $\esym$-node of $\nettwo$.
		\item \emph{Box interiors}: $\bboxp\linktwo\subseteq\nettwo$ for any $\oc$-link $\linktwo \in\nettwo$.
		
		\item \emph{Box free variables}: $\bboxp\link \subseteq \nettwo$ if a free variable of $\bboxp \link$ is internal to $\nettwo$.		
	\end{itemize}
\end{definition}

\paragraph{Decomposing correct nets.} Sequentialisation shall read back an expression by progressively decomposing a correct net. We first need some terminology about boxes.

\begin{definition}[Kinds of boxes]
	Let $\net$ be a correct net. A $\oc$-link $\link$ of $\net$ is:
	\begin{itemize}
		\item \emph{free} if it is at level 0 in $\net$ and its free variables are free variables of $\net$.
		\item an \emph{argument} if its $\esym$-node is the target of a $\otimes$-link;
		\item a \emph{substitution} if its $\esym$-node is the target of a $\set{\weak,\der,\ctxhole}$-link (or, equivalently, if it is not the target of a $\otimes$-link).		
	\end{itemize}
\end{definition}

The following lemma states that, in correct nets whose root structure is similar to the translation of an expression, it is always possible to decompose the net in correct subnets. The lemma does not state the correctness of the interior of boxes because they are correct by definition of correctness.

\begin{toappendix}
\begin{lemma}[Decomposition]
	\label{l:decomposition}
	Let $\net$ be a correct net.
	\begin{enumerate}
	\item \emph{Free weakening}: if $\net$ has a free weakening $\link$ then $\linkset\net \setminus \link$ is a subnet of $\net$.	  
	  \item \emph{Root abstraction}: if the root link $\link$ of $\net$ is a $\parr$-link then $\linkset\net \setminus \link$ is a subnet of $\net$.	  
	  \item \emph{Free substitution}: if $\net$ has a free substitution $\link$ then $\linkset \net \setminus (\set\link \cup \bboxp\link)$ is a subnet of $\net$.
	  \item \emph{Root application with free argument}: if the root link $\link$ of $\net$ is a $\tens$-link whose argument is a free $\oc$-link $\linktwo$ then $\linkset \net \setminus (\set{\link,\linktwo} \cup \bboxp\linktwo)$ is a subnet of $\net$.
	  
	\end{enumerate}
\end{lemma}
\end{toappendix}

\begin{definition}[Decomposable net]
  A correct net $\net$ is decomposable if it is in one of the hypothesis of the decomposition lemma (\reflemma{decomposition}), that is, if it has a free weakening, a root abstraction, a free substitution, or a root application with free argument.
\end{definition}

The last bit is to prove that every correct net is decomposable, and so, essentially corresponds to the translation of an expression.

\begin{toappendix}
\begin{lemma}[Correct nets are decomposable]
\label{l:correct-implies-decomposable}
Let $\net$ be a correct net with more than one link. Then $\net$ is decomposable.
\end{lemma}
\end{toappendix}

We now introduce the read back of correct net as expressions, which is the key notion for the sequentialisation theorem. Its definition relies, in turn, on the various ways in which a correct net can be decomposed, when it has more than one link.

\begin{definition}[Read back]
 Let $\net$ be a correct net and $\expr$ be an expression. The relation \emph{$\expr$ is a read back of $\net$}, noted $\net \readback \expr$, is defined by induction on the number of links in $\net$:
 \begin{itemize}
   \item \emph{One link term net}: $\net$ is a $\der$-link of $\esym$-node $\var$. Then $\net \readback \var$;
   
   \item \emph{One link context net}: $\net$ is a $\ctxhole$-link of $\esym$-nodes $\nameset$. Then $\net \readback \ctxholens$;
   
   \item \emph{Free weakening}: $\net$ has a free weakening $\link$ and $\net \setminus \link \readback \expr$. Then $\net \readback \expr$;
   
   \item \emph{Root abstraction}: the root link $\link$ of $\net$ is a $\parr$-link of $\esym$-node $\var$ and $\net \setminus \link \readback \expr$. Then $\net \readback \la\var\expr$;
   
   \item \emph{Free substitution}: $\net$ has a free substitution $\link$ of $\esym$-node $\var$, $\net \setminus (\set\link \cup \bboxp\link) \readback \expr$, and $\bboxp\link \readback \exprtwo$. Then $\net \readback \expr\esub\var\exprtwo$.
   
   \item \emph{Root application with free argument}: the root link $\link$ of $\net$ is a $\tens$-link whose argument is a free $\oc$-link $\linktwo$, $\net \setminus (\set{\link,\linktwo} \cup \bboxp\linktwo) \readback \expr$, and $\bboxp\linktwo \readback \exprtwo$. Then $\net \readback \expr\exprtwo$.
	  
 \end{itemize}

\end{definition}

We conclude the section with the sequentialisation theorem, that relates terms and proof nets at the static level. Its formulation is slightly stronger than similar theorems in the literature, that usually do not provide completeness.

\begin{toappendix}
\begin{theorem}[Sequentialisation]
\label{thm:sequentialisation} 
Let $\net$ be a correct net and $\varset$ be the set of $\esym$-nodes of its free weakenings. 
\begin{enumerate}
  \item
  \emph{Read backs exist}: there exists $\expr$ such that $\net \readback \expr$ with $\fv\expr = \fv\net$.
  
  \item 
  \emph{The read back relation is correct}: for all expressions $\expr$, $\net \readback \expr$ implies $\lamtonetsvar{\expr}{\varset}=\net$ and $\fv{\net}=\fv{\expr}\cup \varset$.
  
  \item 
  \emph{The read back relation is complete}: if $\lamtonetsvar{\expr}{\namesettwo} = \net$ then $\net \readback \expr$ and $\namesettwo \subseteq \fv\net \cup \nameset$.
\end{enumerate}
\end{theorem}
\end{toappendix}

\begin{figure}[t]
\centering\ovalbox{
\begin{tabular}{\spc c\spc\spc |\spc\spc c\spc\spc |\spc\spc c\spc}
\footnotesize $(\la\vartwo\tm)\esub{\var}{\tmtwo}$  
&
\footnotesize $(\tm\,\tmthree)\esub{\var}{\tmtwo}$  
&
\footnotesize $\tm\esub{\var}{\tmtwo}\esub{\vartwo}{\tmthree}$ 
\\
            $\eqstruct_{\l}$ 
            &
            $\eqstruct_{@l}$
            &
            $\eqstruct_{com}$
            \\
  
            $\la\vartwo\tm\esub{\var}{\tmtwo}$
            &
            $\tm\esub{\var}{\tmtwo}\,\tmthree$
            &
            $\tm\esub{\vartwo}{\tmthree}\esub{\var}{\tmtwo}$
  \\
  
  \footnotesize if $\vartwo \not\in \fv{\tmtwo}$
&
\footnotesize if $\var \not\in \fv{\tmthree}$
&
\footnotesize if $\vartwo \not\in \fv{\tmtwo}$ and $\var \not\in \fv{\tmthree}$ \\

\begin{tikzpicture}[ocenter]
\node at (0,0) [mportn] (fun){};
\node at (fun.center) [mportn, below right=\stalt and \hstlar] (body){};
\inetcell[inductiveTr, at=(body.center), below=1pt](funbody){\footnotesize $t$}[180]
\node at (funbody.left pax) [eportn, below =\sepboxshort, label=below:$\vartwo$] (oc){};

\lparnangle{fun}{oc}{body}{abs}{in=-135, out=180, looseness=1.3, overlay};

\node at (funbody.right pax) [eportn, below =\sepboxshort, label=right:$\var$] (body2){};
\node at (body2.center) [eportn, below =\stalt ] (arginter2){};
\lbangn{arginter2}{body2}{bang2}
\inetcell[inductiveTr, at=(arginter2.center), below=1pt](funbody2){\footnotesize $\tmtwo$}[180]
\node at (funbody2.middle pax) [nospace, below =\sepboxshort] (boxDownSpacing2){};
\aBoxWithPalOnTopAndVeticalNodeBound{bang2}{boxDownSpacing2}{exbox}{15pt}{15pt}
\lbangn{arginter2}{body2}{bang2}
\end{tikzpicture}
&
 \begin{tikzpicture}[ocenter]
\node at (0,0) [mportn] (out){};
\node at (out.center) [nospace, above=7pt] (spacingnode){};
\node at (out.center) [mportn, below left=\stalt and 1.2*\stlar] (fun){};
\node at (out.center) [eportn, below right=\stalt and 1.2*\stlar] (arg){};
\ltensn{out}{fun}{arg}{ap};

\node at (arg.center) [mportn, below =\stalt ] (arginter){};
\lbangn{arginter}{arg}{bang}
\inetcell[inductiveTr,at=(arginter.center), below=1pt](argbody){\footnotesize $\tmthree$}[180]    
\node at (argbody.middle pax) [nospace, below =\sepboxshort] (boxDownSpacing){};
\aBoxWithPalOnTopAndVeticalNodeBound{bang}{boxDownSpacing}{exbox}{15pt}{15pt}
\lbangn{arginter}{arg}{bang}

\inetcell[inductiveTr, at=(fun.center), below=1pt](funbody){\footnotesize $\tm$}[180]

\node at (funbody.right pax) [eportn, below =\sepboxshort, label=right:$\var$] (body2){};
\node at (body2.center) [eportn, below =\stalt ] (arginter2){};
\lbangn{arginter2}{body2}{bang2}
\inetcell[inductiveTr, at=(arginter2.center), below=1pt](funbody2){\footnotesize $\tmtwo$}[180]
\node at (funbody2.middle pax) [nospace, below =\sepboxshort] (boxDownSpacing2){};
\aBoxWithPalOnTopAndVeticalNodeBound{bang2}{boxDownSpacing2}{exbox}{15pt}{15pt}
\lbangn{arginter2}{body2}{bang2}
\end{tikzpicture}

&

 \begin{tikzpicture}[ocenter]
\node at (0,0) [mportn] (fun){};
\inetcell[inductiveTr, at=(fun.center), below=1pt](funbody){\footnotesize $\tm$}[180]

\node at (funbody.right pax) [eportn, below =\sepboxshort, label=right:$\var$] (body2){};
\node at (body2.center) [letic, below right=\hstalt and \hstlar] (bang2){\scriptsize $!$};
\node at (bang2.center) [eportn, below =\hstalt ] (arginter2){};
\draw[eprincn, exppuren, out=-90, in=90] (body2) to (bang2);
\draw[puren] (arginter2) to (bang2);

\inetcell[inductiveTr, at=(arginter2.center), below=1pt](funbody2){\footnotesize $\tmtwo$}[180]
\node at (funbody2.middle pax) [nospace, below =\sepboxshort] (boxDownSpacing2){};
\aBoxWithPalOnTopAndVeticalNodeBound{bang2}{boxDownSpacing2}{exbox}{15pt}{15pt}
\node at (body2.center) [letic, below right=\hstalt and \hstlar] (bang2){\scriptsize $!$};

\node at (funbody.left pax) [eportn, below =\sepboxshort, label=left:$\vartwo$] (leftbox){};
\node at (leftbox.center) [letic, below left=\hstalt and \hstlar] (bang){\scriptsize $!$};
\node at (bang.center) [eportn, below =\hstalt ] (arginter){};
\draw[eprincn, exppuren, out=-90, in=90] (leftbox) to (bang);
\draw[puren] (arginter) to (bang);

\inetcell[inductiveTr,at=(arginter.center), below=1pt](argbody){\footnotesize $\tmthree$}[180]    
\node at (argbody.middle pax) [nospace, below =\sepboxshort] (boxDownSpacing){};
\aBoxWithPalOnTopAndVeticalNodeBound{bang}{boxDownSpacing}{exbox}{15pt}{15pt}
\node at (leftbox.center) [letic, below left=\hstalt and \hstlar] (bang){\scriptsize $!$};

\end{tikzpicture}

\end{tabular}
}
\caption{\label{fig:quotient} Structural equivalent terms translate to the same proof nets (contractions of common variables are omitted).}
\end{figure}
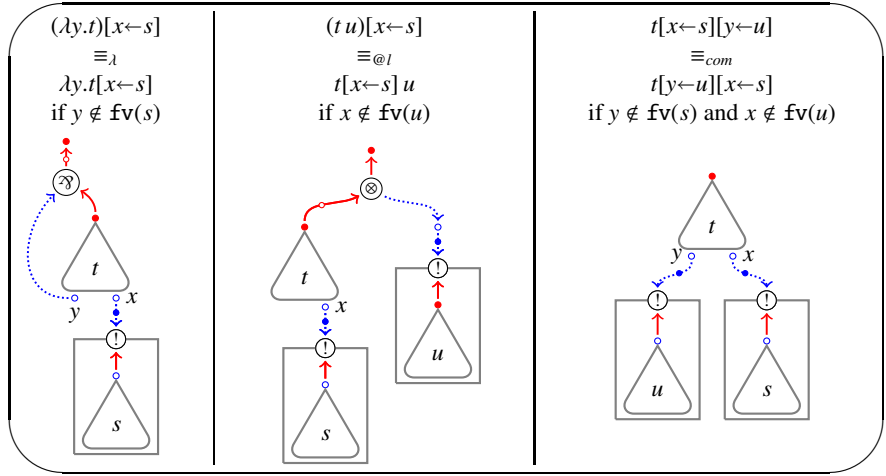

\paragraph{Quotient.} Next we prove that structural equivalence on the LSC is exactly the quotient induced by proof nets. We invite the reader to look at the proof of the following quotient theorem. The $\Leftarrow$ direction essentially follows from figure \reffig{quotient}, where for simplicity we have omitted the contractions of common variables for the subnets. The $\Rightarrow$ direction is the tricky point. Note that $\eqstruct$-classes do not admit canonical representantives, because the $\eqstruct_{com}$ axiom is not orientable, and so it is not possible to rely on some canonical read back. The argument at work in the proof is however pleasantly simple.

\begin{toappendix}
\begin{theorem}[Quotient]
\label{thm:quotient}
Let $\net$ be correct term net. Then, $\lamtonets\tm = \net$ and $\lamtonets\tmtwo = \net$ if and only if  $\tm \eqstruct \tmtwo$.  
\end{theorem}
\end{toappendix}

\section{Contexts}

This short section develops a few notions about relating contexts in the two frameworks. We only deal with  what is strictly needed to relate rewriting steps on terms and on term nets---a more general treatment is possible, but not explored here, for the sake of simplicity.

The plugging operation can also be done on context nets. 

\begin{definition}[Plugging on context nets]
Let $\net$ be a context net and let $\nameset$ be the free variables of its $\ctxhole$-link $\link$. The \emph{plugging} of a net $\nettwo$ with free variables $\namesettwo\subseteq \nameset$ in $\net$ is the net $\net\ctxholep\nettwo$ obtained by 
\begin{itemize} 
  \item if $\link$ is at level 0:
\begin{itemize} 
  \item \emph{Replacement}: replacing $\link$ with $\nettwo$;
  \item \emph{Weakening unused variables in the interface}: adding a weakening $\linktwo$ on every variable $\var \in (\nameset \setminus \namesettwo)$ not shared in $\net$ (or whose only incoming link in $\net$ is $\link$).
\end{itemize}
  \item if $\link$ is in $\bboxp\linktwo$ for a $\oc$-link $\linktwo$ at level 0 then:  
  \begin{itemize} 
  \item \emph{Recursive plugging}: replacing the links of $\bboxp\linktwo$ with those in $\bboxp\linktwo\ctxholep\nettwo$, inheriting the boxes;  
  \item \emph{Pushing weakenings out of the box}: redefining $\bboxp\linktwo$ as $\bboxp\linktwo\ctxholep\nettwo$ less its free weakenings, if any.
\end{itemize}  
\end{itemize}
\end{definition}

The next lemma relies plugging in context nets with the corresponding read backs.

\begin{toappendix}
\begin{lemma}[Properties of context nets plugging]
\label{l:context-net-plugging}
  Let $\net$ be a context net of interface $\nameset$, $\nettwo$ a correct net with free variables $\namesettwo\subseteq \nameset$. Then 
  \begin{enumerate}
    \item \emph{Correctness}: $\net\ctxholep\nettwo$ is correct;
    \item \emph{Read back}: if $\net \readback \ctxns$ and $\nettwo \readback \expr$ then $\net\ctxholep\nettwo \readback \ctxnsp\expr$.
  \end{enumerate}
\end{lemma}
\end{toappendix}

From the read back property, a dual property follows for the translation.

\begin{toappendix}
\begin{lemma}[Context-free translation]
\label{l:context-free}
  Let $\ctxns$ a context, $\expr$ an expression such that $\fv\expr \subseteq \nameset$, and $\namesettwo$ a set of variables. Then $\lamtonetsvar{\ctxnsp\expr}{\namesetthree} = \lamtonetsvar{\ctxns}{\namesettwo}\ctxholep{\lamtonets{\expr}}$ where $\namesetthree = \namesettwo \cup (\nameset \setminus \cv\ctxns)$.
\end{lemma}
\end{toappendix}

The following lemma shall be used to relate the exponential steps in the two systems. The proof is a straightforward but tedious induction on $\net \readback \ctxns$, which is omitted.

\begin{lemma}[Read back and free variable occurrences]
\label{l:readback-factorisation}  
Let $\net \readback \tm$ be a term net with a fixed read back, $\link$ be a $\der$-link of $\net$ whose $\esym$-node $\var$ is a free variable of $\net$. Then for every set of variable names $\nameset$ there are a context $\ctx$ and a context net $\nettwo$, both of interface $\nameset\cup\set\var$, such that 
\begin{enumerate}
\item \emph{Net factorisation}: $\nettwo\ctxholep\link = \net$;
\item \emph{Term factorisation}: $\ctxfp\var = \tm$; and
\item \emph{Read back}: $\nettwo \readback \ctx$.
\end{enumerate}
\end{lemma}

%

%
%


\section{Micro-Step Operational Semantics}
\label{sect:pn-dynamics}
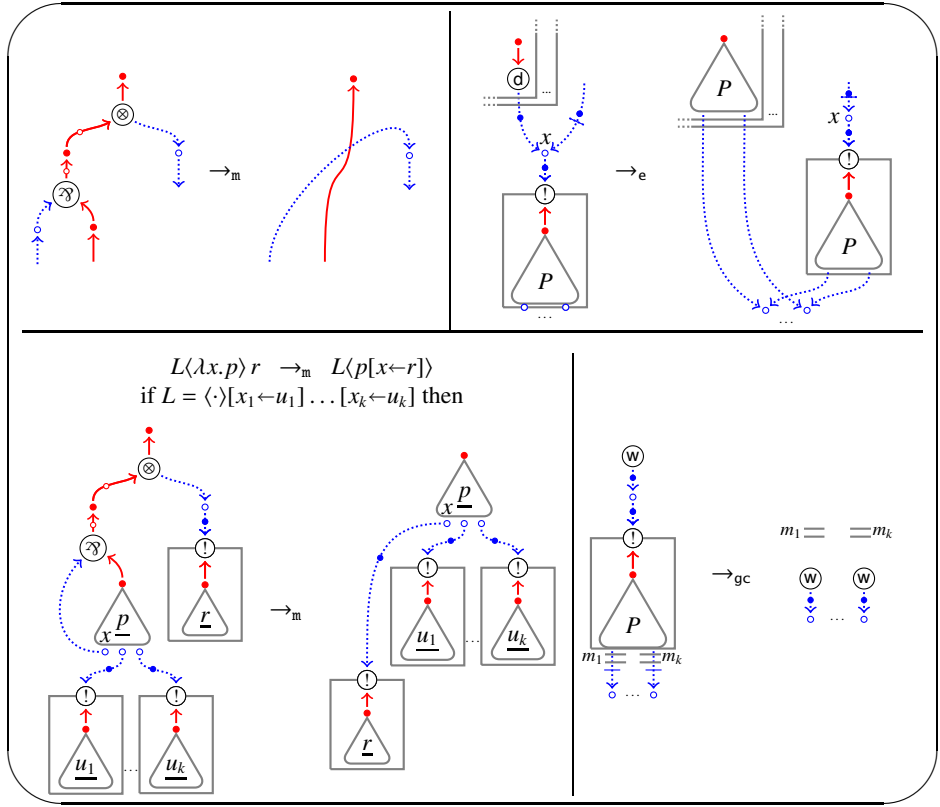
\begin{figure}[t] 
\centering\ovalbox{
\begin{tabular}{cccc}
\begin{tabular}{c\spc |\spc ccc}
\begin{tabular}{c\spc c\spc c}
\begin{tikzpicture}[ocenter]
\node [mportn] (out){};
\node at (out.center) [mportn, below left=\stalt and \stlar] (fun){};
\node at (out.center) [eportn, below right=\stalt and \stlar] (arg){};
\ltensn{out}{fun}{arg}{ap};

\node at (arg.center)[nospace, below=\hstalt](ghost){};
\draw[exppuren] (arg) to (ghost);

\node at (fun.center) [below right=\stalt and \hstlar, mportn](downr){};
\node at (fun.center) [eportn, below left=\stalt and \hstlar](x){};
\lparn{fun}{x}{downr}{par}

\node at (downr.center)[nospace, below=\hstalt](ghostparm){};
\draw[puren] (ghostparm) to (downr);

\node at (x.center)[nospace, below=\hstalt](ghostpare){};
\draw[exppuren] (ghostpare) to (x);

\end{tikzpicture}
&
$\tom$
&

\begin{tikzpicture}[ocenter]
\node at (0,0)[mportn](out){};
\node at (out.center)[eportn, below right=\stalt and \stlar](tense){};

\node at (out.center)[nospace, below left=2.5*\stalt and \hstlar](ghostparm){};
\draw[puren, out=90, in=-90,looseness=2] (ghostparm) to (out);

\node at (ghostparm.center)[nospace, left=\stlar](ghostpare){};
\draw[exppuren, out=90, in=90,looseness=1] (ghostpare) to (tense);

\node at (arg.center)[nospace, below=\hstalt](ghost){};
\draw[exppuren] (arg) to (ghost);

\end{tikzpicture}
\end{tabular}
&
\begin{tabular}{c\spc c\spc c}
\begin{tikzpicture}[ocenter]
\node at (0,0) [mportn] (o1){};
\node at (o1.center) [below= \hstalt,fill=white, letic] (der){\scriptsize $\der$};
\draw[puren] (o1) to (der);
\node at (o1.center) [below right=1.5*\stalt and \hstlar, label=above: $\var$, eportn] (input){};
\draw[eprincn, exppuren, out=-89,in=150] (der) to (input);
\node at (der.center)[below right = .8*\altax and .5*\ilar, nospace](boxes){};
\boxcornersright{boxes}{.7*\stlar}{1.7*\hstalt}{exboxline}

\node at (der.center) [nospace, right =1.2*\stlar] (ghostocr){};
\draw[eprincn,genexppuren, out=-90, in=30](ghostocr)to(input);

\node at (input.center) [mportn, below =\stalt] (beta){};
\lbangn{beta}{input}{bang}

\inetcell[inductiveTr,at=(beta.center), below=1pt](argbody){}[180]
\node at (argbody.center) [nospace] (z1){\footnotesize $\net$};
\node at (argbody.left pax) [below=\sepboxshort,eportn] (z1){};
\node at (argbody.right pax) [below=\sepboxshort,eportn] (zn){};
\gdots{z1}{zn}{below=.5pt}

\aBoxWithPalOnTopAndVeticalNodeBound{bang}{z1}{exbox}{15pt}{15pt}
\lbangn{beta}{input}{bang}
\node at (argbody.left pax) [below=\sepboxshort,eportn] (z1){};
\node at (argbody.right pax) [below=\sepboxshort,eportn] (zn){};



\end{tikzpicture}
&
$\toe$
&
\begin{tikzpicture}[ocenter]
\node at (0,0) [mportn] (bodyn){};
\inetcell[inductiveTr, at=(bodyn.center), below=1pt](subbody){\footnotesize $\net$}[180]

\node at (subbody.right pax)[below right = 2pt and .5*\ilar, nospace](boxes){};
\boxcornersright{boxes}{1.3*\stlar}{1.2*\stalt}{exboxline}

\node at (bodyn) [below right = \hstalt and 2.3*\stlar, nospace] (ghostocl){};
\node at (ghostocl.center) [nospace, right =\stlar] (ghostocr){};
\node at (ghostocl.center) [eportn, below =\hstalt, label=left:$\var$] (oc){};

\draw[eprincn,genexppuren](ghostocl)to(oc);

\node at (oc.center) [mportn, below =\stalt] (bangint){};
\lbangn{bangint}{oc}{bang}

\inetcell[inductiveTr, at=(bangint.center), below=1pt](subbody2){\footnotesize $\net$}[180]
\node at (subbody2.middle pax) [nospace, below =\sepboxshort] (oc2){};
\aBoxWithPalOnTopAndVeticalNodeBound{bang}{oc2}{exbox}{15pt}{15pt}
\lbangn{bangint}{oc}{bang}

\node at (subbody.left pax|-subbody2.left pax) [nospace] (dummyl){};
\node at (subbody.right pax|-subbody2.right pax) [nospace] (dummyr){};

\node at \med{dummyl}{subbody2.left pax} [eportn, below =\hstalt] (oc3){};
\node at \med{dummyr}{subbody2.right pax} [eportn, below =\hstalt] (oc4){};
\gdots{oc3}{oc4}{below=1pt}


\draw[exppuren, in=150, out=-90](subbody.right pax)to(oc4);
\draw[exppuren, in=45, out=-90](subbody2.right pax)to(oc4);
\draw[exppuren, in=150, out=-90](subbody.left pax)to(oc3);
\draw[exppuren, in=45, out=-90](subbody2.left pax)to(oc3);

\end{tikzpicture}
\end{tabular}
\end{tabular}
\\[1.7cm]\hline\\[-.1cm]
\begin{tabular}{c | ccc}

\begin{tabular}{c}
$\sctxp{\la\var\tmfive}\, \tmfour\ \ \tom\ \ \sctxp{\tmfive\esub\var\tmfour}$\\ if $\sctx = \ctxhole\esub{\var_1}{\tmthree_1}\ldots\esub{\var_k}{\tmthree_k}$ then \\[.2cm]

\begin{tabular}{c\spc c\spc  c}
 \begin{tikzpicture}[ocenter]
\node [mportn] (out){};
\node at (out.center) [mportn, below left=\stalt and \stlar] (fun){};
\node at (out.center) [eportn, below right=\stalt and \stlar] (arg){};
\ltensn{out}{fun}{arg}{ap};
\node at (arg.center) [mportn, below =\stalt ] (arginter){};
\lbangn{arginter}{arg}{argBang}
\inetcell[inductiveTrSmall,at=(arginter.center), below=1pt](argbody){\footnotesize $\lamtonets\tmfour$}[180]    

\node at (fun.center) [mportn, below right=\stalt and \hstlar] (body){};
\inetcell[inductiveTrSmall, at=(body.center), below=1pt](funbody){\footnotesize $\lamtonets\tmfive$}[180]
\node at (funbody.left pax) [eportn, below =\sepboxshort, label=above:$\var$] (oc){};

\lparnangle{fun}{oc}{body}{abs}{in=-135, out=180, looseness=1.3, overlay};

\node at (argbody.middle pax) [nospace, below =\sepboxshort] (argBoxDownSpacing){};
\aBoxWithPalOnTopAndVeticalNodeBound{argBang}{argBoxDownSpacing}{exbox}{13pt}{13pt}
\lbangn{arginter}{arg}{argBang}

\node at (funbody.right pax) [eportn, below =\sepboxshort] (body2){};
\node at (body2.center) [letic, below right=\hstalt and 1.1*\hstlar] (bang2){\scriptsize $!$};
\node at (bang2.center) [mportn, below =.8*\hstalt ] (arginter2){};
\draw[eprincn, exppuren, out=-90, in=90] (body2) to (bang2);
\draw[puren] (arginter2) to (bang2);

\inetcell[inductiveTrSmall, at=(arginter2.center), below=1pt](funbody2){\footnotesize $\lamtonets{\tmthree_k}$}[180]
\node at (funbody2.middle pax) [nospace, below =\sepboxshort] (boxDownSpacing2){};
\aBoxWithPalOnTopAndVeticalNodeBound{bang2}{boxDownSpacing2}{exbox}{13pt}{13pt}
\node at (body2.center) [letic, below right=\hstalt and 1.1*\hstlar] (bang2){\scriptsize $!$};

\node at (funbody.middle pax) [eportn, below =\sepboxshort] (leftbox){};
\node at (leftbox.center) [letic, below left=\hstalt and 1.1*\hstlar] (bang){\scriptsize $!$};
\node at (bang.center) [mportn, below =.8*\hstalt ] (boxlinter){};
\draw[eprincn, exppuren, out=-90, in=90] (leftbox) to (bang);
\draw[puren] (boxlinter) to (bang);

\inetcell[inductiveTrSmall,at=(boxlinter.center), below=1pt](argbody2){\footnotesize $\lamtonets{\tmthree_1}$}[180]    
\node at (argbody2.middle pax) [nospace, below =\sepboxshort] (boxDownSpacing){};
\aBoxWithPalOnTopAndVeticalNodeBound{bang}{boxDownSpacing}{exbox}{13pt}{13pt}
\node at (leftbox.center) [letic, below left=\hstalt and 1.1*\hstlar] (bang){\scriptsize $!$};

\gdots{funbody2}{argbody2}{}
\end{tikzpicture}

&
$\tom$
&

 \begin{tikzpicture}[ocenter]

\node at (fun.center) [mportn, below right=\stalt and \hstlar] (body){};
\inetcell[inductiveTrSmall, at=(body.center), below=1pt](funbody){\footnotesize $\lamtonets\tmfive$}[180]
\node at (funbody.left pax) [eportn, below =\sepboxshort, label=above:$\var$] (oc){};

\node at (funbody.right pax) [eportn, below =\sepboxshort] (body2){};
\node at (body2.center) [letic, below right=\hstalt and 1.1*\hstlar] (bang2){\scriptsize $!$};
\node at (bang2.center) [mportn, below =.8*\hstalt ] (arginter2){};
\draw[eprincn, exppuren, out=-90, in=90] (body2) to (bang2);
\draw[puren] (arginter2) to (bang2);

\inetcell[inductiveTrSmall, at=(arginter2.center), below=1pt](funbody2){\footnotesize $\lamtonets{\tmthree_k}$}[180]
\node at (funbody2.middle pax) [nospace, below =\sepboxshort] (boxDownSpacing2){};
\aBoxWithPalOnTopAndVeticalNodeBound{bang2}{boxDownSpacing2}{exbox}{13pt}{13pt}
\node at (body2.center) [letic, below right=\hstalt and 1.1*\hstlar] (bang2){\scriptsize $!$};

\node at (funbody.middle pax) [eportn, below =\sepboxshort] (leftbox){};
\node at (leftbox.center) [letic, below left=\hstalt and 1.1*\hstlar] (bang){\scriptsize $!$};
\node at (bang.center) [mportn, below =.8*\hstalt ] (boxlinter){};
\draw[eprincn, exppuren, out=-90, in=90] (leftbox) to (bang);
\draw[puren] (boxlinter) to (bang);

\inetcell[inductiveTrSmall,at=(boxlinter.center), below=1pt](argbody2){\footnotesize $\lamtonets{\tmthree_1}$}[180]    
\node at (argbody2.middle pax) [nospace, below =\sepboxshort] (boxDownSpacing){};
\aBoxWithPalOnTopAndVeticalNodeBound{bang}{boxDownSpacing}{exbox}{13pt}{13pt}
\node at (leftbox.center) [letic, below left=\hstalt and 1.1*\hstlar] (bang){\scriptsize $!$};

\gdots{funbody2}{argbody2}{}

\node at (boxDownSpacing.center) [letic, below left=.2*\hstalt and 2*\hstlar] (newbang){\scriptsize $!$};
\node at (newbang.center) [mportn, below =.8*\hstalt ] (newboxlinter){};
\draw[eprincn, exppuren, out=180, in=90, looseness=1.2] (oc) to (newbang);
\draw[puren] (newboxlinter) to (newbang);

\inetcell[inductiveTrSmall,at=(newboxlinter.center), below=1pt](newbody){\footnotesize $\lamtonets\tmfour$}[180]    
\node at (newbody.middle pax) [nospace, below =\sepboxshort] (newboxDownSpacing){};
\aBoxWithPalOnTopAndVeticalNodeBound{newbang}{newboxDownSpacing}{exbox}{13pt}{13pt}
\node at (boxDownSpacing.center) [letic, below left=.2*\hstalt and 2*\hstlar] (newbang){\scriptsize $!$};
\end{tikzpicture}\end{tabular}
\end{tabular}
&
\begin{tabular}{c\spc c\spc c}
\begin{tikzpicture}[ocenter]
\node at (0,0) [eportn] (input){};
\lweakn{input}{weakening};
\node at (input.center) [mportn, below =\stalt] (beta){};
\lbangn{beta}{input}{bang}

\inetcell[inductiveTr,at=(beta.center), below=1pt](argbody){}[180]
\node at (argbody.center) [nospace] (z1){\footnotesize $\net$};
\node at (argbody.left pax) [below=3.5*\sepbox,eportn] (z1){};
\node at (argbody.right pax) [below=3.5*\sepbox,eportn] (zn){};
\gdots{z1}{zn}{}

\aBoxWithPalOnTopAndVeticalNodeBound{bang}{argbody.left pax}{exbox}{15pt}{15pt}
\lbangn{beta}{input}{bang}

\draw[genexppuren, exppuren](argbody.right pax) to (zn);
\draw[genexppuren, exppuren](argbody.left pax) to (z1);

\node at (argbody.left pax) [below left=2pt and 0.1*\stlar](inlub){};
\node at (argbody.left pax) [below right=2pt and 0.1*\stlar](inrub){};
\node at (argbody.left pax) [below left=-1pt and 0.1*\stlar](inldb){};
\node at (argbody.left pax) [below right=-1pt and 0.1*\stlar](inrdb){};
\draw[exboxline](inlub.center)--(inrub.center);
\draw[exboxline](inldb.center)--(inrdb.center);
\node at \med{inlub}{inldb}[left=2pt, etic](m1){\scriptsize $m_1$};

\node at (argbody.right pax) [below left=2pt and 0.1*\stlar](inlubk){};
\node at (argbody.right pax) [below right=2pt and 0.1*\stlar](inrubk){};
\node at (argbody.right pax) [below left=-1pt and 0.1*\stlar](inldbk){};
\node at (argbody.right pax) [below right=-1pt and 0.1*\stlar](inrdbk){};
\draw[exboxline](inlubk.center)--(inrubk.center);
\draw[exboxline](inldbk.center)--(inrdbk.center);
\node at \med{inrubk}{inrdbk}[right=2pt, etic](mk){\scriptsize $m_k$};
\end{tikzpicture}
&
$\togc$
&
\begin{tikzpicture}[ocenter]
\node at (0,0) [eportn] (input){};
\lweakn{input}{weakening};

\node at (input.center) [right=\stlar, eportn] (input2){};
\lweakn{input2}{weakening2};

\gdots{input}{input2}{}

\node at (input) [above left=\stalt+3pt and 0.1*\stlar](inlub){};
\node at (input) [above right=\stalt+3pt and 0.1*\stlar](inrub){};
\node at (input) [above left=\stalt and 0.1*\stlar](inldb){};
\node at (input) [above right=\stalt and 0.1*\stlar](inrdb){};
\draw[exboxline](inlub.center)--(inrub.center);
\draw[exboxline](inldb.center)--(inrdb.center);
\node at \med{inlub}{inldb}[left=2pt, etic](m1){\scriptsize $m_1$};

\node at (input2) [above left=\stalt+3pt and 0.1*\stlar](inlubk){};
\node at (input2) [above right=\stalt+3pt and 0.1*\stlar](inrubk){};
\node at (input2) [above left=\stalt and 0.1*\stlar](inldbk){};
\node at (input2) [above right=\stalt and 0.1*\stlar](inrdbk){};
\draw[exboxline](inlubk.center)--(inrubk.center);
\draw[exboxline](inldbk.center)--(inrdbk.center);
\node at \med{inrubk}{inrdbk}[right=2pt, etic](mk){\scriptsize $m_k$};

\end{tikzpicture}
\end{tabular}
\end{tabular}
\end{tabular}
}
\caption{Proof nets cut-elimination rules, plus---in the bottom-left corner---the matching of the multiplicative rule on terms and on term nets (forgetting, for simplicity, about the contraction of common variables for the boxes, and the fact that $\var_j$ can occur in $\tmthree_i$ for $i<j$).\label{fig:rew-rules}}
\end{figure}
Here we define the rewriting rules on proof nets and prove the isomorphism of rewriting systems with respect to the LSC. Since the rules of the LSC and those of proof nets match perfectly, we use the same names and the same notations for them.

\paragraph{The rules.} The rewriting rules are in \reffig{rew-rules}. Let us explain them. First of all, note that the notion of cut in our syntax is implicit, because cut-links are not represented explicitly. A cut is given by a node whose incoming and outgoing connections are principal (\ie\ with a little dot on the line). 

The multiplicative rule $\tom$ is nothing but the usual elimination of a multiplicative cut, adapted to our syntax. The matching with the rule on terms is shown in \reffig{rew-rules}.

The garbage collection rule $\togc$ corresponds to a cut with a weakening. It is mostly as the usual rule, the only difference is with respect to the reduct. The box of the $\oc$-link is erased and replaced by a set of weakenings, one for every free variable of $\nettwo$---this is standard. Each one of these new weakenings is also pushed out of all the $m_i$ boxes closing on its $\esym$-node. This is done to preserve the invariant that weakenings are always pushed out of boxes as much as possible. Such an invariant is also used  in the rule: note that the weakening is at the same level of $\nettwo$. Last, if the weakenings created by the rule are contracted with any other link then they are removed on the fly, because by definition weakenings cannot be contracted.

The Milner exponential rule $\toe$ is the most unusual rule, and---to our knowledge---it has never been considered before on proof nets. There are two unusual points about it. The first one is that the redex crosses box borders, as the $\der$-link is potentially inside many boxes, while the $\oc$-link is out of those boxes. In the literature, this kind of rules is usually paired with a small-step operational semantics (\eg in \cite{Reg:Thesis:92}), that is, all the copies of the box are done in a single shot. Here instead we employ a micro-step semantics, as also done in \cite{DBLP:conf/rta/Accattoli13}---that paper contains a discussion about this \emph{box-crossing principle} and its impact on the rewriting theory of proof nets.

The second unusual point is the way the cut is eliminated. Roughly, it corresponds to a duplication of the box (so a contraction cut-elimination) immediately followed by commutation with all the boxes and opening of the box (so a dereliction cut-elimination). We say \emph{roughly}, because there is a difference: the duplication happens also if the $\der$-link is not contracted. Exactly as in the LSC, indeed, the $\toe$ rule duplicates the ES even if there are no other occurrences of the replaced variable. In case the $\der$-link is not contracted, the rule puts a weakening on the $\esym$-node source of the $\oc$-link.

\paragraph{The isomorphism.} Finally, we relate the evaluation of proof nets and of the LSC.

\begin{toappendix}
\begin{theorem}[Dynamic isomorphism]
\label{thm:dynamic-isomorphism}
Let $\net \readback \tm$ be a correct net with a fixed read back, and $a\in\set{\msym,\esym,\gcsym}$. There is a bijection $\phi$ between $a$-redexes of $\tm$ and $\net $ such that:
\begin{enumerate}
\item \emph{Terms to proof nets}: given a redex $\redex: \tm \Rew{a} \tmtwo$ then there exists $\nettwo$ such that $\phi(\redex): \net \Rew{a} \nettwo$ and $\nettwo \readback \tmtwo$.
\item \emph{Proof nets to terms}: given a redex $\redex: \net \Rew{a} \nettwo$  then there exists $\tmtwo$ such that $\phi^{-1}(\redex):\tm \Rew{a} \tmtwo$ and $\nettwo \readback \tmtwo$.
\end{enumerate}
\end{theorem}
\end{toappendix}
From \refthm{dynamic-isomorphism} it immediately follows that cut-elimination preserves correctness, because the reduct of a correct net is the translation of a term, and therefore it is correct.

\begin{corollary}[Preservation of correctness]
Let $\net$ be a term net and $\net\Rew{} \nettwo$. Then $\nettwo$ is correct.
\end{corollary}

The perfect matching also transfers to proof nets the residual system of the LSC defined in \cite{DBLP:conf/popl/AccattoliBKL14}.
Finally, the dynamic isomorphism (\refthm{dynamic-isomorphism}) combined with the quotient theorem (\refthm{quotient}) also provides a new proof of the strong bisimulation property of structural equivalence (\refprop{bisimulation}).
\section{Abstracting Proof Nets From a Rewriting Point of View}
In this section we provide a new, rewriting-based perspective on proof nets.

\paragraph{Cut commutes with cut.} One of the motivations for proof nets is the fact that cut-elimination in the sequent calculus has to face commutative cut-elimination cases. They are always a burden, but most of them are harmless. There is however at least one very delicate case, the commutation of cut with itself, given by:
\begin{center}
  \begin{tabular}{c\spc c\spc ccc}
		\indproof{$\vdash \Gamma,  \red\formtwo$}{$\gamma$}
		\indproof{$\vdash\Gamma,  \red\form$}{$\pi$}
		\indproof{$\vdash \Gamma, \red\form, \red\formtwo$}{$\theta$}

		\RightLabel{cut}
		\BinaryInfC{$\vdash \Gamma,  \red\formtwo$}

		\RightLabel{cut}
		\BinaryInfC{$\vdash\Gamma$}			\DisplayProof 
&
  $\to$
  &
		\indproof{$\vdash\Gamma,  \red\form$}{$\pi$}
		\indproof{$\vdash \Gamma,  \red\formtwo$}{$\gamma$}
		\indproof{$\vdash \Gamma, \red\form, \red\formtwo$}{$\theta$}

		\RightLabel{cut}
		\BinaryInfC{$\vdash \Gamma,  \red\form$}

		\RightLabel{cut}
		\BinaryInfC{$\vdash\Gamma$}			
		\DisplayProof 
		\end{tabular}
\end{center}
Such a commutation is delicate because it can be iterated, creating silly loops. If one studies weak normalisation (\ie the \emph{existence} of a normalising path) then it is enough to design a cut-elimination strategy that never commutes cut with itself---this is what is done in the vast majority of cut-elimination theorems. But if one is interested in strong normalisation (\ie, \emph{all} paths eventually normalise), then this is a serious issue. Morally, this is the conceptual problem behind proof nets and also behind the design of good explicit substitution calculi---it could be said that it is \emph{the} rewriting issue of the Curry-Howard correspondence at the micro-step granularity.

One way to address this problem is to introduce an equivalence relation $\sim$ on proofs including the commutation of cut with itself, and then to switch to eliminate cuts modulo $\sim$. Rewriting modulo is a studied but technical and subtle topic, see \cite{Terese} chapter 14.3. The problem is that cut-elimination $\to$ and $\sim$ in general do not interact nicely, in particular $\sim$ cannot be postponed, because it \emph{creates} $\to$-redexes.

Proof nets are a different, more radical solution: a change of syntax in which $\sim$-classes collapse on a single object, the proof net, so that the problem of the interaction between $\to$ and $\sim$ disappears. Proof nets seem, at first, elegant objects, and certainly a brilliant solution to the problem, providing many new intuitions about proofs. They are however heavy to manipulate formally, and it would be often preferable to have an alternative, more traditional syntax with similar properties.

\paragraph{Structural rewriting systems.} The LSC is the prototype of a finer solution to the problem of commuting cut with itself. In general, we said, $\to$ and $\sim$ do not interact nicely. However, it is sometimes possible to \emph{redefine} $\to$ so as to interact nicely with $\sim$. Typically, the contextual rules of the LSC interact nicely with $\equiv$ ($\equiv$ is the equivalence $\sim$ of the LSC, note in particular that axiom $\eqstruct_{com}$ is exactly commutation of cut with itself)---this is the motivation behind contextual rules, sometimes also called \emph{at a distance}. This suggests the following notion, which is a special case of rewriting modulo an equivalence relation.

\begin{definition}[Structural rewriting system]
Let $T$ be a set of objects, $\to$ a rewriting relation and $\sim$ an equivalence relation over $T$. The triple $(T,\to,\sim)$ is a structural rewriting system (modulo) if $\sim$ is a strong bisimulation with respect to $\to$.
\end{definition}

Note that the definition does not mention graphs. We can then see proof nets and the LSC as instances of a single concept. 

\begin{proposition}
let $\Rew{PN}$ be the union of rules $\tom$, $\toe$, and $\togc$ on proof nets.
  \begin{enumerate}
    \item Proof nets with $\Rew{PN}$ are a structural rewriting sytem, by taking $\sim$ to be the identity.
    \item The LSC with $\tolsc$ and $\equiv$ is a structural rewriting sytem. 
  \end{enumerate}  
\end{proposition}

Structural rewriting sytems can be exported to different settings, with no need to bother about correctness criteria or graphical presentations, or the existence of a logical interpretation. For instance, in \cite{DBLP:journals/corr/abs-1302-6337} there is a structural presentation of a fragment of the $\pi$-calculus based on contextual rules, independently of any logical interpretation.

\section{Conclusions}

This paper provides a perfect matching between the LSC and a certain presentation of the fragment of linear logic representing the $\l$-calculus. In particular, we prove that proof nets can be identified with the LSC up to structural equivalence $\eqstruct$, enabling one to reason about proof nets by means of a non-graphical language. 

We also discuss our approach with respect to the basic proof theoretical problem of the cut rule commuting with itself. We try to suggest that the idea behind our result goes beyond proof nets and the LSC, as it also applies to other settings where rewriting has to interact with a notion of structural equivalence such as the $\pi$-calculus.
 
\paragraph{Acknowledgments.} To the reviewers, for useful comments.
This work has been partially funded by the ANR JCJC grant COCA HOLA (ANR-16-CE40-004-01).

\newpage
\phantomsection
\addcontentsline{toc}{section}{References}


%

\newpage
  \appendix
  

\section{Proof Appendix}

\subsection{Proof Nets}

\gettoappendix {l:linear-subnet}

\begin{proof}
Simply note that the length-1-paths between $\msym$-nodes are induced by $\set{\parr,\tens}$-links, and that 
\begin{enumerate}
  \item \emph{Linear shape}: one such node cannot have two imcoming connections, because of the multiplicative node condition on the definition of nets;
  \item \emph{One terminal node}: there is only one terminal $\msym$-node because of the root correctness condition;
  \item \emph{No cycles}: there are no cycles because of the acyclicity correctness condition.
\end{enumerate} 
\end{proof}

\subsection{Correct Nets and Read Backs}

\gettoappendix {l:decomposition}
\begin{proof}
  \hfill
	\begin{enumerate}
	\item \emph{Free weakening}: correctness of $\linkset\net \setminus \link$ is straightforward because the removal of $\link$ cannot affect any of the correctness conditions. The closure conditions for subnets are also trivially true: the target $\esym$-node of $\link$ is not an internal node nor the free variable of any box, because $\link$ is by hypothesis a \emph{free} link.
	  \item \emph{Root abstraction}: the correctness conditions of $\linkset\net \setminus \link$ essentially follow from those for $\net$:
	  \begin{itemize}
	    \item \emph{Root}: the root node of $\net$ is removed with $\link$, but there is a new root $\msym$-node, the source of $\link$.
	    \item \emph{Acyclicity}: removing a link cannot create cycles.
	    \item \emph{Recursive correctness}: by the root condition for $\net$, the root link $\link$ is at level 0 and then out of all boxes, so the removal of $\link$ cannot affect this condition. 
	  \end{itemize}
	  The closure conditions are also true, because the removal creates a new free variables, without changing the set of links on any internal $\esym$-node, nor boxes.

	  \item \emph{Free substitution}: let $\nettwo \defeq \linkset \net \setminus (\set\link \cup \bboxp\link)$. Note that $\zeronet\nettwo$ is simply $\zeronet\net$ without some edges and potentially without some free variables. The correctness conditions of essentially follow from those for $\net$:
	  \begin{itemize}
	    \item \emph{Root}: the root $\msym$-node of $\net$ is the root $\msym$-node of $\nettwo$, as satisfies the root condition in $\zeronet\nettwo$ because it does in $\zeronet\net$---the removal affects only $\esym$-nodes.
	    \item \emph{Acyclicity}: removing edges and nodes cannot create cycles.
	    \item \emph{Recursive correctness}: all boxes of $\nettwo$ are boxes of $\net$, because of the nesting condition for nets. 
	  \end{itemize}
	  The closure conditions are also true, because the removal impacts only on free variables, that is, not on any internal $\esym$-node.
	  
	  \item \emph{Root application with free argument}: let $\nettwo \defeq \linkset \net \setminus (\set{\link,\linktwo} \cup \bboxp\linktwo)$. Correctness of $\nettwo$:
	  \begin{itemize}
	    \item \emph{Root}: the root node of $\net$ is removed with $\link$, but there is a new root $\msym$-node, the $\msym$-source of $\link$.
	    \item \emph{Acyclicity}: removing links cannot create cycles.
	    \item \emph{Recursive correctness}: by the root condition for $\net$, the root link $\link$ is at level 0 and then out of all boxes, so the removal of $\link$ cannot affect this condition. By the nesting condition, removing a free argument cannot affect other boxes. So every box of $\nettwo$ is a box of $\net$.
	  \end{itemize}
	  The closure conditions are also true, because the removal impacts only on free variables, that is, not on any internal $\esym$-node.  
	\end{enumerate}
\end{proof}

\gettoappendix {l:correct-implies-decomposable}
\begin{proof}
If $\net$ has a free weakening or a root abstraction the statement holds. Then assume that it has no free weakenings nor a root abstraction. Consider the following order on the set $S$ of $\oc$-links at level 0: $\link \leq \linktwo$ if there is a path from the $\esym$-node of $\link$ to the $\esym$-node of $\linktwo$ in $\zeronet{\net}$. Acyclicity of $\zeronet{\net}$ implies that $S$ contains maximal elements with respect to $\leq$, if it is non-empty. Two cases:
\begin{itemize}
  \item \emph{$S$ is empty}: then there are no $\oc$-links in $\net$ (see the paragraph about the level of links at page \pageref{def:level}), that implies that there also are no $\tens$-links (see the \emph{tensors and boxes} paragraph at page \pageref{def:arguments}). Then there can only be $\der$-links, and only one of them, otherwise the \emph{root} correctness condition for $\net$ would not hold---absurd, because by hypothesis $\net$ has more than one link.
  
  \item \emph{$S$ is non-empty}: then consider a maximal $\oc$-link $\link$ and suppose that it is not free. By maximality, there cannot be a substitution on one of its free variables. Then, one of its free variables is the $\esym$-node of a $\parr$-link $\linktwo$. Note that all the links on the red path from $\linktwo$ to the root (given by \reflemma{linear-subnet}) are $\parr$-links because 1) there cannot be a $\tens$-link, otherwise $\link$ would have a path to its argument $\linkthree$ and either $\link$ would not be maximal in $S$ (if $\linkthree \neq \link$), against hypothesis, or there would be a cycle in $\zeronet\net$ (if $\linkthree = \link$), against correctness. Then the root link is a $\parr$-link, which is absurd---therefore $\link$ is free. 
  
  Now, if the root link is a $\der$-link then there are no $\tens$-links at level 0 by the linear skeleton lemma (\reflemma{linear-subnet}) and so all elements of $S$ are substitutions. Since $S$ non-empty, its maximal elements are free substitutions and the statement holds.
  
  If the root link is a $\tens$-link then consider its argument $\link$. Note that if $\link$ has a path in $\zeronet\net$ to another argument at level 0 then it has a downward path ending on a $\parr$-link at level 0. But then such a $\parr$-link, being at level 0, has a path to the root link (again by the by the linear skeleton lemma) and so to $\link$, closing a cycle in $\zeronet\net$, against correctness---then $\link$ cannot have paths in $\zeronet\net$ to arguments at level 0. Then either it is maximal in $S$ itself, and then we have a root $\tens$-link with a free argument, as required, or there is a maximal $\oc$-link $\linktwo$ such that $\link \leq \linktwo$, that is a substitution (because it cannot ba an arugment) and it is free (by maximality).
\end{itemize}
\end{proof}

\gettoappendix {thm:sequentialisation}
\begin{proof}
\hfill
\begin{enumerate}
  \item By induction on the number of links of $\net$. By the root and free variables conditions the minimum number of links is 1, and the link must be a $\der$-link. Then $\net \readback \var$ for some $\var$. If $\net$ has more than one link then it is decomposable by \reflemma{correct-implies-decomposable}, and so it decomposes according to the decomposition lemma (\reflemma{decomposition}). We can then apply the \ih and the definition of read back, obtaining $\net \readback \tm$ for some $\tm$.
  
  \item By a straightforward induction on the number of links of $\net$ and case analysis of $\net \readback \tm$.
  
  \item By induction on $\lamtonetsvar{\expr}{\namesettwo} = \net$. Note that when $\expr$ is a variable or a context hole and $\nameset$ is contained in their free variables then $\net$ has only one link and we directly have $\net \readback \expr$. Otherwise, just note that all cases of the translation produce a decomposable net, where the decomposition concerns the topmost constructor of $\expr$. Then the statement follows from the \ih and the definition of read back.
\end{enumerate}

\end{proof}

\gettoappendix {thm:quotient}
\begin{proof}\hfill
  \begin{itemize}
    \item[] $\Rightarrow$) By induction on $\net$. If the topmost constructor of $\tm$ and $\tmtwo$ coincide (in the case of ES they have to subsitute on the same variable) then we decompose the net, apply the \ih, and obtain the statement by recomposing the equivalences of the subterms. If instead the topmost constructors are different, we do one case, the others follow all the same pattern. 
    
  Suppose that $\tm = \la\var\tm'$ and $\tmtwo = \tmtwo'\esub\vartwo\tmthree$. Then $\net$ is decomposable in two different ways: it has a root abstraction binding $\var$ and a free substitution on $\vartwo$---note that, being a \emph{free} substitution, $\var \notin \fv\tmthree$. Let $\net_{\tm'}$ be $\net$ without the root abstraction---we have $\lamtonets{\tm'} = \net_{\tm'} $. It still has a free substitution, so that among the possible read backs of $\net_{\tm'}$ we have $\net_{\tm'} \readback \tm''\esub\vartwo\tmthree$ for some term $\tm''$ such that $\nettwo \readback \tm''$ where $\nettwo$ is the subnet of $\net_{\tm'}$ obtained by removing the free substitution. By correctness of read back (\refthm{sequentialisation}.2), $\lamtonets{\tm''\esub\vartwo\tmthree} = \net_{\tm'}$, and so by \ih $\tm'\eqstruct \tm''\esub\vartwo\tmthree$. 
    
    Repeating the reasoning by first decomposing with respect to $\tmtwo$ and the free substitution, we obtain a net $\net_{\tmtwo'} \readback \la\var\tmtwo''$ for some term $\tmtwo''$ such that $\tmtwo' \eqstruct \la\var\tmtwo''$, and a subnet $\netthree$ of $\net_{\tmtwo'}$ such that $\netthree \readback \tmtwo''$. 
    
    Note that $\nettwo = \netthree$, because both are obtained by removing the root abstraction and the free abstraction from $\net$, and the two operations commute. Then $\nettwo \readback \tm''$ and $\nettwo \readback \tmtwo''$. By correctness of read back  $\lamtonets{\tm''} = \nettwo$ and $\lamtonets{\tmtwo''} = \nettwo$, and so by \ih, $\tm''\eqstruct \tmtwo''$. Now,
    $$\tm = \la\var\tm' \eqstruct_{\ih} \la\var(\tm''\esub\vartwo\tmthree) \eqstruct (\la\var\tm'')\esub\vartwo\tmthree
    \eqstruct_{\ih} (\la\var\tmtwo'')\esub\vartwo\tmthree \eqstruct_{\ih}  \tmtwo'\esub\vartwo\tmthree = \tmtwo$$
    
    \item[] $\Leftarrow$) By induction on $\tm \eqstruct \tmtwo$. The base cases are in \reffig{quotient} at page \pageref{fig:quotient}. The contextual and trasitive closure follow by the \ih
  \end{itemize}

\end{proof}

\subsection{Contexts}

\gettoappendix {l:context-net-plugging}
\begin{proof}\hfill
  \begin{enumerate}
    \item By induction on the level $k$ of the $\ctxhole$-link $\link$. Cases:
    \begin{itemize}
      \item $k=0$. Correctness conditions:
      \begin{itemize}
      \item \emph{Root}: about the $\msym$-nodes of $\zeronet{\net\ctxholep\nettwo}$ coming from $\zeronet\net$, the replacement preserves the terminal one and does not turn any other $\msym$-node into a terminal one. About the $\msym$-nodes of $\zeronet{\net\ctxholep\nettwo}$ coming from $\zeronet\nettwo$, the replacement preserves the terminal one only if it coincides with the one of $\zeronet\net$ (if $\ctxhole$-link was the root link of $\net$) and does not turn any other $\msym$-node into a terminal one. Then the condition is satisfied.
      \item \emph{Acyclicity}: the acyclic structure of $\link$ is replaced by the acyclic structure $\zeronet\nettwo$. Note that in both cases the nodes on the boundary are targets of the structure, so changing the internal structure cannot create cycles.
      \item \emph{Recursive correctness}: the internal of boxes of $\net\ctxholep\nettwo$ are those of $\net$ plus those of $\nettwo$, which are all unaffected by the replacement---so the condition follows from the one for $\net$.
      \end{itemize}
      \item $k>0$. The difference between $\zeronet\net$ and $\zeronet{\net\ctxholep\nettwo}$ amounts to some free weakenings, if any. Then, the root and acyclicity conditions for $\zeronet{\net\ctxholep\nettwo}$  follows from those for $\zeronet\net$. Recursive correctness for all boxes at level 0 also follows from the one for $\zeronet\net$ but for the one containing the $\ctxhole$-link, for which it follows by the \ih      
    \end{itemize}
    
    \item By a straightforward but tedious induction on $\net \readback \ctxns$.
  \end{enumerate}  
\end{proof}

\gettoappendix {l:context-free}
\begin{proof}
  Consider $\lamtonetsvar{\ctxns}{\namesetthree}$ and $\lamtonets{\expr}$, that by \refprop{transl-is-correct} are a context and a term net. By \reflemma{context-net-plugging}, $\lamtonetsvar{\ctxns}{\namesettwo}\ctxholep{\lamtonets{\expr}}$ is correct and reads back to $\ctxnsp\expr$. By correctness of sequentialisation (\refthm{sequentialisation}.2), there is a set of variable $\namesetthree$ such that $\lamtonetsvar{\ctxnsp\expr}{\namesetthree} = \lamtonetsvar{\ctxns}{\namesettwo}\ctxholep{\lamtonets{\expr}}$. Last, note that the plugging $\lamtonetsvar{\ctxns}{\namesettwo}\ctxholep{\lamtonets{\expr}}$  may introduce some free weakening beyond those potentially introduced by $\namesettwo$, given by names in $\nameset$ that are not captured by $\ctxns$ nor appear free in $\ctxns$ or $\expr$.
\end{proof}

\subsection{Operational Semantics}

\gettoappendix {thm:dynamic-isomorphism}

\begin{figure}[t]
\begin{center}
\ovalbox{

\begin{tabular}{c}
$\sctxp{\la\var\tmfive}\, \tmfour\ \ \tom\ \ \sctxp{\tmfive\esub\var\tmfour}$\\ if $\sctx = \ctxhole\esub{\var_1}{\tmthree_1}\ldots\esub{\var_k}{\tmthree_k}$ then \\[.2cm]

\begin{tabular}{c\spc c\spc  c}
 \begin{tikzpicture}[ocenter]
\node [mportn] (out){};
\node at (out.center) [mportn, below left=\stalt and \stlar] (fun){};
\node at (out.center) [eportn, below right=\stalt and \stlar] (arg){};
\ltensn{out}{fun}{arg}{ap};
\node at (arg.center) [mportn, below =\stalt ] (arginter){};
\lbangn{arginter}{arg}{argBang}
\inetcell[inductiveTrSmall,at=(arginter.center), below=1pt](argbody){\footnotesize $\lamtonets\tmfour$}[180]    

\node at (fun.center) [mportn, below right=\stalt and \hstlar] (body){};
\inetcell[inductiveTrSmall, at=(body.center), below=1pt](funbody){\footnotesize $\lamtonets\tmfive$}[180]
\node at (funbody.left pax) [eportn, below =\sepboxshort, label=above:$\var$] (oc){};

\lparnangle{fun}{oc}{body}{abs}{in=-135, out=180, looseness=1.3, overlay};

\node at (argbody.middle pax) [nospace, below =\sepboxshort] (argBoxDownSpacing){};
\aBoxWithPalOnTopAndVeticalNodeBound{argBang}{argBoxDownSpacing}{exbox}{13pt}{13pt}
\lbangn{arginter}{arg}{argBang}

\node at (funbody.right pax) [eportn, below =\sepboxshort] (body2){};
\node at (body2.center) [letic, below right=\hstalt and 1.1*\hstlar] (bang2){\scriptsize $!$};
\node at (bang2.center) [mportn, below =.8*\hstalt ] (arginter2){};
\draw[eprincn, exppuren, out=-90, in=90] (body2) to (bang2);
\draw[puren] (arginter2) to (bang2);

\inetcell[inductiveTrSmall, at=(arginter2.center), below=1pt](funbody2){\footnotesize $\lamtonets{\tmthree_k}$}[180]
\node at (funbody2.middle pax) [nospace, below =\sepboxshort] (boxDownSpacing2){};
\aBoxWithPalOnTopAndVeticalNodeBound{bang2}{boxDownSpacing2}{exbox}{13pt}{13pt}
\node at (body2.center) [letic, below right=\hstalt and 1.1*\hstlar] (bang2){\scriptsize $!$};

\node at (funbody.middle pax) [eportn, below =\sepboxshort] (leftbox){};
\node at (leftbox.center) [letic, below left=\hstalt and 1.1*\hstlar] (bang){\scriptsize $!$};
\node at (bang.center) [mportn, below =.8*\hstalt ] (boxlinter){};
\draw[eprincn, exppuren, out=-90, in=90] (leftbox) to (bang);
\draw[puren] (boxlinter) to (bang);

\inetcell[inductiveTrSmall,at=(boxlinter.center), below=1pt](argbody2){\footnotesize $\lamtonets{\tmthree_1}$}[180]    
\node at (argbody2.middle pax) [nospace, below =\sepboxshort] (boxDownSpacing){};
\aBoxWithPalOnTopAndVeticalNodeBound{bang}{boxDownSpacing}{exbox}{13pt}{13pt}
\node at (leftbox.center) [letic, below left=\hstalt and 1.1*\hstlar] (bang){\scriptsize $!$};

\gdots{funbody2}{argbody2}{}
\end{tikzpicture}

&
$\tom$
&

 \begin{tikzpicture}[ocenter]

\node at (fun.center) [mportn, below right=\stalt and \hstlar] (body){};
\inetcell[inductiveTrSmall, at=(body.center), below=1pt](funbody){\footnotesize $\lamtonets\tmfive$}[180]
\node at (funbody.left pax) [eportn, below =\sepboxshort, label=above:$\var$] (oc){};

\node at (funbody.right pax) [eportn, below =\sepboxshort] (body2){};
\node at (body2.center) [letic, below right=\hstalt and 1.1*\hstlar] (bang2){\scriptsize $!$};
\node at (bang2.center) [mportn, below =.8*\hstalt ] (arginter2){};
\draw[eprincn, exppuren, out=-90, in=90] (body2) to (bang2);
\draw[puren] (arginter2) to (bang2);

\inetcell[inductiveTrSmall, at=(arginter2.center), below=1pt](funbody2){\footnotesize $\lamtonets{\tmthree_k}$}[180]
\node at (funbody2.middle pax) [nospace, below =\sepboxshort] (boxDownSpacing2){};
\aBoxWithPalOnTopAndVeticalNodeBound{bang2}{boxDownSpacing2}{exbox}{13pt}{13pt}
\node at (body2.center) [letic, below right=\hstalt and 1.1*\hstlar] (bang2){\scriptsize $!$};

\node at (funbody.middle pax) [eportn, below =\sepboxshort] (leftbox){};
\node at (leftbox.center) [letic, below left=\hstalt and 1.1*\hstlar] (bang){\scriptsize $!$};
\node at (bang.center) [mportn, below =.8*\hstalt ] (boxlinter){};
\draw[eprincn, exppuren, out=-90, in=90] (leftbox) to (bang);
\draw[puren] (boxlinter) to (bang);

\inetcell[inductiveTrSmall,at=(boxlinter.center), below=1pt](argbody2){\footnotesize $\lamtonets{\tmthree_1}$}[180]    
\node at (argbody2.middle pax) [nospace, below =\sepboxshort] (boxDownSpacing){};
\aBoxWithPalOnTopAndVeticalNodeBound{bang}{boxDownSpacing}{exbox}{13pt}{13pt}
\node at (leftbox.center) [letic, below left=\hstalt and 1.1*\hstlar] (bang){\scriptsize $!$};

\gdots{funbody2}{argbody2}{}

\node at (boxDownSpacing.center) [letic, below left=.2*\hstalt and 2*\hstlar] (newbang){\scriptsize $!$};
\node at (newbang.center) [mportn, below =.8*\hstalt ] (newboxlinter){};
\draw[eprincn, exppuren, out=180, in=90, looseness=1.2] (oc) to (newbang);
\draw[puren] (newboxlinter) to (newbang);

\inetcell[inductiveTrSmall,at=(newboxlinter.center), below=1pt](newbody){\footnotesize $\lamtonets\tmfour$}[180]    
\node at (newbody.middle pax) [nospace, below =\sepboxshort] (newboxDownSpacing){};
\aBoxWithPalOnTopAndVeticalNodeBound{newbang}{newboxDownSpacing}{exbox}{13pt}{13pt}
\node at (boxDownSpacing.center) [letic, below left=.2*\hstalt and 2*\hstlar] (newbang){\scriptsize $!$};
\end{tikzpicture}\end{tabular}
\end{tabular}
}
\end{center}
\caption{\label{fig:multiplicative-matching}Matching of the multiplicative rule on terms and on term nets (forgetting, for simplicity, about the contraction of common variables for the boxes, and the fact that $\var_j$ can occur in $\tmthree_i$ for $i<j$), referred to from the proof of \refthm{dynamic-isomorphism}.}
\end{figure}
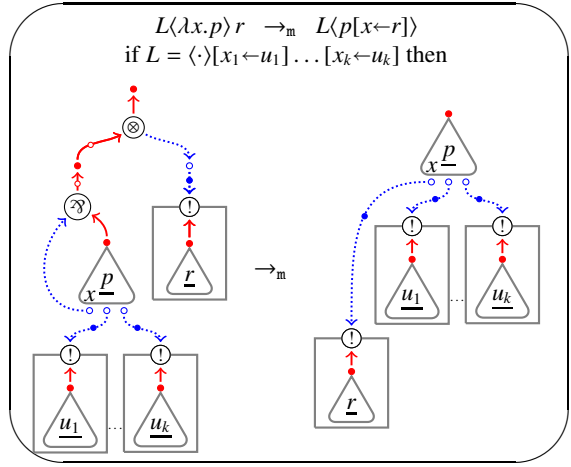

\begin{proof}
By induction on $\net \readback \tm$. Cases:
\begin{itemize}
\item \emph{Variable}, \ie $\tm = \var$. Then it has no redexes. The net $\net$ is then a $\der$-link plus possibly some weakenings, and it also has no redexes.

\item \emph{Abstraction}, \ie $\tm = \la\var\tmthree$. Then the root link $\link$ of $\net$ is a $\parr$-link and $\net \setminus \link \readback \tmthree$. By \ih, there is a bijection $\phi$ satisfying the statement between $\tmthree$ and $\net \setminus \link$. The same bijection works for $\tm = \la\var\tmthree$ and $\net$, because adding the abstraction and the $\parr$-link does not add redexes. The reducts according to these redexes are evidently still in the read back relation.

\item \emph{Application}, \ie $\tm = \tmthree \tmfour$. Then by definition of read back the root link $\link$ of $\net$ is a $\tens$-link whose argument is a free $\oc$-link $\linktwo$, $\netthree \defeq \net \setminus (\set{\link,\linktwo} \cup \bboxp\linktwo) \readback \tmthree$, and $\bboxp\linktwo \readback \tmfour$. By \ih, there are bijections of redexes satisfying the statement between $\netthree$ and $\tmthree$, and between $\bboxp\linktwo$ and $\tmfour$. All these redexes are redexes of $\tm$ and $\net$ that are then in bijection. Moreover, the reducts according to these redexes are evidently still in the read back relation.
\begin{enumerate}
\item If there is a redex $\redex$ in $\tm$ that is not contained in $\tmthree$ nor $\tmfour$ it necessarily involves the root application, and so it is a $\tom$ redex. Then $\tmthree$ has the form $\sctxp{\la\var\tmfive}$ and so the root link of $\netthree$ is a $\parr$-link, that is, $\net$ has a $\tom$-redex not contained in $\netthree$ nor $\bboxp\linktwo$---we set $\phi(\redex)$ to be this redex. Now, simply note that the term reduct $\tmthree$ translates to the net reduct $\nettwo$ as in \reffig{multiplicative-matching}, that is, $\lamtonetsvar\tmthree\nameset = \nettwo$ for some $\nameset$, and so $\nettwo \readback \tmtwo$ by completeness of read back (\refthm{sequentialisation}.3).

\item If there is a redex $\redex$ in $\net$ that is not contained in $\netthree$ nor $\bboxp\linktwo$ it necessarily involves the root $\tens$-link, and so it is a $\tom$ redex. Then $\netthree$ has a root $\parr$-link, and so $\tmthree$ has the form $\sctxp{\la\var\tmfive}$ for some substitution context $\sctx$, that is, $\tm$ has a $\tom$-redex not contained in $\tmthree$ nor $\tmfour$---we set $\phi^{-1}(\redex)$ to be this redex. Again, simply note that the term reduct $\tmthree$ translates to the net reduct $\nettwo$ as in \reffig{multiplicative-matching}, that is, $\lamtonetsvar\tmthree\nameset = \nettwo$ for some $\nameset$, and so $\nettwo \readback \tmtwo$ by completeness of read back (\refthm{sequentialisation}.3).
\end{enumerate}

\item \emph{Substitution}, \ie $\tm = \tmthree \esub\var\tmfour$. By definition of read back, $\net$ has a free substitution $\link$ of $\esym$-node $\var$, $\netthree \defeq \net \setminus (\set\link \cup \bboxp\link) \readback \tmthree$, and $\bboxp\link \readback \tmfour$. By \ih, there are bijections of redexes satisfying the statement between $\netthree$ and $\tmthree$, and between $\bboxp\linktwo$ and $\tmfour$. All these redexes are redexes of $\tm$ and $\net$ that are then in bijection. 

The redexes of $\tm$ and $\net$ not contained in their subterms / subnets, then, have to involve the variable $\var$ and the substitution on it. By \refprop{transl-is-correct}, the variable $\var$ has the same multiplicity in both $\tmthree$ and $\netthree$. Then if $\tm$ has a $\tow$ redex on $\var$ so does $\net$, and viceversa. And if $\tm$ has $n$ $\toe$ redexes on $\var$ so does $\net$, and viceversa. The bijection is then extablished.

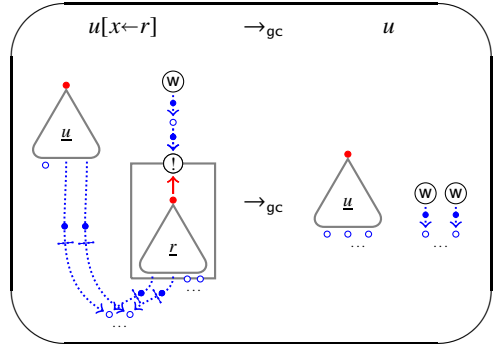
\begin{figure}[t]
\begin{center}
\ovalbox{
\begin{tabular}{c\spc c\spc c}
$\tmthree \esub\var\tmfour$ & $\togc$ & $\tmthree$\\\\
\begin{tikzpicture}[ocenter]
\node at (0,0) [mportn] (bodyn){};
\inetcell[inductiveTr, at=(bodyn.center), below=1pt](subbody){\scriptsize $\lamtonets\tmthree$}[180]
\node at (subbody.left pax) [eportn, below =\sepboxshort] (var){};

\node at (bodyn) [below right = \hstalt and 2*\stlar, eportn] (weaknode){};
\lweakn{weaknode}{weakening}

\node at (weaknode.center) [mportn, below =\stalt] (bangint){};
\lbangn{bangint}{weaknode}{bang}

\inetcell[inductiveTr, at=(bangint.center), below=1pt](subbody2){\scriptsize $\lamtonets\tmfour$}[180]
\node at (subbody2.middle pax) [nospace, below =\sepboxshort] (oc2){};
\aBoxWithPalOnTopAndVeticalNodeBound{bang}{oc2}{exbox}{15pt}{15pt}
\lbangn{bangint}{weaknode}{bang}

\node at (subbody.middle pax|-subbody2.left pax) [nospace] (dummyl){};
\node at (subbody.right pax|-subbody2.right pax) [nospace] (dummyr){};

\node at \med{dummyl}{subbody2.left pax} [eportn, below =\hstalt] (oc3){};
\node at \med{dummyr}{subbody2.middle pax} [eportn, below =\hstalt] (oc4){};
\gdots{oc3}{oc4}{below=1pt}


\draw[genexppuren, eprincn, in=150, out=-90](subbody.right pax)to(oc4);
\draw[genexppuren, eprincn, in=45, out=-90](subbody2.middle pax)to(oc4);
\draw[genexppuren, eprincn, in=150, out=-90](subbody.middle pax)to(oc3);
\draw[genexppuren, eprincn, in=45, out=-90](subbody2.left pax)to(oc3);

\node at (subbody2.right pax) [eportn, below left=\sepboxshort and 1.5*\sepboxshort] (vartoweak){};
\node at (subbody2.right pax) [eportn, below right=\sepboxshort and 1.5*\sepboxshort] (vartoweak2){};
\gdots{vartoweak}{vartoweak2}{below = 1pt}
\end{tikzpicture}
&
$\togc$
&
\begin{tikzpicture}[ocenter]
\node at (0,0) [mportn] (bodyn){};
\inetcell[inductiveTr, at=(bodyn.center), below=1pt](subbody){\scriptsize $\lamtonets\tmthree$}[180]
\node at (subbody.left pax) [eportn, below =\sepboxshort] (var){};
\node at (subbody.middle pax) [eportn, below =\sepboxshort] (var2){};
\node at (subbody.right pax) [eportn, below =\sepboxshort] (var3){};

\gdots{var2}{var3}{below = 1pt}

\node at (var3.center) [eportn, right =\stlar] (var4){};
\lweakn{var4}{weakening}

\node at (var4.center) [eportn, right =\hstlar] (var5){};
\lweakn{var5}{weakening2}
\gdots{var4}{var5}{below = 1pt}
\end{tikzpicture}
\end{tabular}
}
\end{center}
\caption{\label{fig:weakening-matching}Matching of the garbage collection rule on terms and on term nets, referred to from the proof of \refthm{dynamic-isomorphism}.}
\end{figure}

The matching of the reducts for $\tow$ is given by \reffig{weakening-matching}.

For $\toe$:
\begin{enumerate}
  \item Let $\redex: \tm = \ctxfp\var \esub\var\tmfour \toe \ctxfp\tmfour \esub\var\tmfour = \tmtwo$. We need to spell out some notation. We can assume that the interface of $\ctx$ contains $\var$ and $\fv\tmfour$, that is, that $\tmthree = \ctxnp{\fv\tmfour\cup\set\var}\var$. Let also $\nameset$ be the set of variables such that $\lamtonetsvar{\tmthree}{\nameset}  = \lamtonetsvar{\ctxfp\var}{\nameset} = \netthree$. Finally, by definition of net the box $\bboxp\link$ has no free weakening, so $\lamtonets\tmfour = {\bboxp\link}$. 

Now, we can reason. By context-freeness of the translation (\reflemma{context-free}), $\netthree = \lamtonetsvar{\ctxnp{\fv\tmfour\cup\set\var}\var}{\nameset} = \lamtonetsvar{\ctxn{\fv\tmfour\cup\set\var}}{\nameset}\ctxholep{\lamtonets{\var}}$ and $\lamtonetsvar{\ctxnp{\fv\tmfour\cup\set\var}\tmfour}{\nameset} = \lamtonetsvar{\ctxn{\fv\tmfour\cup\set\var}}{\nameset}\ctxholep{\lamtonets{\tmfour}} =\lamtonetsvar{\ctx}{\nameset}\ctxholep{\lamtonets{\bboxp\link}}$. Then the translation of $\tmtwo$ is given by $\lamtonetsvar{\ctx}{\nameset}\ctxholep{\lamtonets{\bboxp\link}}\cup\set\link \cup \bboxp\link$ that is exactly the $\toe$ reduct of $\net$.

  \item Let $\redex: \net \toe \nettwo$ and let $\linkthree$ be the $\der$-link of $\netthree$ of $\esym$-node $\var$ substituted by the rewriting step. By the factorisation property of read backs (\reflemma{readback-factorisation}), there are a context $\ctx$, a context net $\netfour$, both of interface $\fv\tmfour\cup\set\var$ and such that $\netfour\ctxholep\linkthree = \netthree$, $\ctxp\var = \tmthree$, and $\netfour \readback \ctx$. Then, the redex $\net$ is equal to $\netthree \cup \set\link \cup \bboxp\link$ and the read back associated to it has the shape $\tm = \ctxfp\var \esub\var\tmfour$. On the other hand, the reduct $\nettwo$ is equal to $\netfour\ctxholep{\bboxp\link}\cup \set\link \cup \bboxp\link$ and, by context-freeness of the translation (\reflemma{context-free}), the term $\tmtwo \defeq \ctxfp\tmfour \esub\var\tmfour$ translates to it, namely there is a set of variable names $\nameset$ such that $\lamtonetsvar\tmtwo\nameset = \nettwo$. By completeness of the read back relation (\refthm{sequentialisation}.3), $\nettwo \readback \tmtwo$. And, of course, $\tm = \ctxfp\var \esub\var\tmfour \toe \ctxfp\tmfour \esub\var\tmfour = \tmtwo$.
\end{enumerate}
\end{itemize}
\end{proof}

\end{document}